\documentclass[12pt,english,longbibliography,nofootinbib,superscriptaddress,12pt,sort&compress,showkeys]{article}
\usepackage{ae,aecompl}
\usepackage[T1]{fontenc}
\usepackage{textcomp}
\usepackage[latin9]{inputenc}
\usepackage[active]{srcltx}
\usepackage{float}
\usepackage{amsmath}
\usepackage{graphicx}
\usepackage[numbers]{natbib}

\makeatletter

\providecommand{\tabularnewline}{\\}

\newcommand{\lyxaddress}[1]{
	\par {\raggedright #1
	\vspace{1.4em}
	\noindent\par}
}

\usepackage{aecompl}\usepackage{epstopdf}

\usepackage[raggedrightboxes]{ragged2e}

\usepackage{babel}

\makeatother

\usepackage{babel}
\begin{document}
\title{A model of full thermodynamic stabilization of nanocrystalline alloys}
\author{Omar Hussein and Yuri Mishin}
\maketitle

\lyxaddress{Department of Physics and Astronomy, MSN 3F3, George Mason University,
Fairfax, Virginia 22030, USA}
\begin{abstract}
\noindent We propose a model of a polycrystalline alloy combining
the Potts model for grain orientations with a lattice-gas model for
solute thermodynamics and diffusion. The alloy evolution with this
model is implemented by kinetic Monte Carlo simulations with nonlinear
transition barriers between microstates. The model is applied to investigate
the long-standing question of whether grain boundary (GB) segregation
of an appropriate solute can drive the GB free energy to zero, creating
a fully stabilized polycrystalline state with a finite grain size.
The model reproduces stable polycrystalline states under certain conditions,
provided the solute-solute interactions are repulsive. The material's
structure minimizing the total free energy is not static. It exists
in a state of dynamic equilibrium between the competing processes
of grain growth and grain refinement. The alloy eliminates triple
junctions by forming a set of smaller grains embedded into a larger
matrix grain. It is predicted that, if a fully stabilized nanocrystalline
state is realized experimentally, it will look very different from
the conventional (unstable) nanocrystalline materials. 
\end{abstract}

\section{Introduction\label{sec:Intro}}

Over the past decades, nanocrystalline (NC) materials have attracted
much attention in both research and technological applications \citep{MeyersMB06,Darling:2026aa}.
Some mechanical and physical properties of NC materials are superior
to those of their conventional, coarse-grained counterparts. However,
NC materials are thermodynamically unstable because grain boundaries
(GBs) possess an excess free energy relative to the perfect crystalline
lattice. This excess free energy causes extensive grain growth at
elevated temperatures. Once the grains grow in size well above the
nanoscale, the superior properties of the NC material are lost.

Significant research efforts have been dedicated to finding ways to
stabilize NC materials against grain growth by alloying with suitable
solutes. Two ways to achieve stabilization have been pursued: decreasing
the GB free energy $\gamma$ by GB segregation of appropriate solutes,
or reducing the GB mobility. These two strategies are often referred
to as the thermodynamic and kinetic stabilization mechanisms. 

The basic principle behind the thermodynamic stabilization can be
traced back to Gibbs \citep{Gibbs}. The Gibbs adsorption equation
predicts a reciprocal relation between interface segregation and the
reduction in the interface tension. In fact, it is this reduction
in free energy that drives the interface segregation. There has been
extensive research, both experimental and theoretical, aiming to find
solutes that cause significant reduction in the GB free energy. The
search was aided by constructing nano-stability diagrams based on
thermodynamic calculations for many alloy systems \citep{Detor2007,Chookajorn2012,Chookajorn2014,Kalidindi:2017aa,Kalidindi:2017bb,Kalidindi:2017cc,Murdoch:2013aa,Trelewicz2009,Perrin-2021,Koch2008,Darling2014,Saber2013,Saber2013a,Xing:2018aa,Zhou:2014aa}.
Experimentally, grain growth retardation by alloying has been demonstrated
in numerous alloys, although the extent of the retardation varies.
When significant stabilization was achieved, it was not always clear
if the success could be attributed to a reduction in $\gamma$ or
suppression of GB mobility. One of the challenges in this approach
is that one can only pack that much solute in GBs. After that, the
solute atoms start precipitating as a bulk phase. 

In the kinetic mechanism, the GB mobility is reduced by the solute
drag effect \citep{Abdeljawad:2015aa,Abdeljawad:2017aa,Alkayyali:2021uo,Cahn-1962,Wang03,Toda-Caraballo2013,Koju:2020aa,Hillert:1999aa,Hillert:2001aa,Mishin:2019aa,Koju:2021aa,Suhane:2022aa,Svoboda:2011aa,Mishin:2023aa,Mishin:2023ab}
or Zener pinning of GBs by small particles \citep{Abdeljawad:2017aa,Li:1990aa,Koju:2016aa,Manohar:1998aa,Miodownik:2000aa,Nes-1985,Di-Prinzio:2013aa,Zhou:2017ab,Koju:2018ub,Hornbuckle:2015aa}.
A prominent example of extremely strong nano-stabilization is offered
by Cu-Ta alloys \citep{Koju:2016aa,Hornbuckle:2021aa,Darling:2016aa,Kale:2023aa,Koju:2018ub,Hornbuckle:2015aa}.
Ta is virtually immiscible in Cu. When a small amount of Ta is forced
into Cu by mechanical alloying followed by consolidation and thermal
processing, Ta precipitates from the metastable solution as nanoclusters
coherent with the Cu matrix. These nanoclusters pin the GBs in place
and prevent grain growth. The grains do not grow significantly up
to 0.9 of the melting point. This high-temperature NC alloy exhibits
extraordinary properties, such as high strength of about a GPa, excellent
creep resistance, fatigue endurance, and even shock deformation resistance.
Thus, in practical terms, the Zener pinning of GBs is probably the
most effective nano-stabilization mechanism known today. Although
other approaches, such as solute drag and lowering the GB free energy
by segregation, are less effective, they nevertheless present significant
fundamental interest.

In the 1990s, Weissmuller \citep{Weissmuller1994,Weissmuller:1993aa}
and later others \citep{Kirchheim:2002aa,Liu:2004aa,Kirchheim2007a,Kirchheim2007b,Krill:2005aa,Schvindlerman2006,Detor2007,Chookajorn2012,Chookajorn2014,Kalidindi:2017aa,Kalidindi:2017bb,Kalidindi:2017cc,Murdoch:2013aa,Trelewicz2009,Perrin-2021,NC-stability}
discussed the possibility of decreasing the GB free energy until the
total free energy of the polycrystal reaches a minimum at a \emph{finite}
grain size.\footnote{More precisely, the free energy $F$ is minimized with respect to
the fraction $f$ of GB sites: $dF/df=0$. It is assumed that all
GBs have the same width $\delta$ and uniform properties. Assuming
also that $f\propto\delta/D$, where $D>0$ is the average grain size,
the stability condition becomes $dF/dD=0$. } In the minimum-free-energy state, the driving force for grain growth
vanishes, and the NC material becomes \emph{fully stabilized}. Thermodynamic
analysis within the uniform boundary model predicts that in order
to achieve the fully stabilized state, the GB free energy $\gamma$
must converge to a zero value when the system arrives at the equilibrium
state \citep{Hussein:2024aa}. 

This fascinating idea has been debated in the literature for many
years. At first sight, it may seem that it contradicts Gibbs' thermodynamics.
Gibbs predicted that as long as $\gamma\leq0$, the material will
keep increasing its interface area by breaking into infinitely thin
lamellas. However, this thought experiment assumes an open system
with unlimited supply of solute atoms to the interfaces. Real nano-grains
have a finite capacity to supply the solute to GBs. At some degree
of grain refinement, the grains will no longer be able to provide
enough solute to sustain the $\gamma\leq0$ condition. At this point,
the process of grain refinement should stop, and the polycrystalline
system should reach equilibrium between GB segregation and the solid
solution in the grains. 

To our knowledge, a fully stabilized state has never been demonstrated
experimentally. In addition to the mentioned competition between GB
segregation and bulk phase transformations, it might be difficult
to distinguish thermodynamic stabilization from kinetic trapping on
the experimental time scales. There are also many unanswered fundamental
questions surrounding the full stabilization. For example, GB segregation
is known to depend on the GB crystallography. Can the $\gamma=0$
condition be achieved simultaneously at all GBs in a polycrystalline
sample? Even if it can, which is unlikely, the triple junction (TJ)
free energy does not have to approach zero when GBs do. If the TJ
free energy remains positive, a driving force for structural evolution
remains, and the polycrystal is not fully stabilized. One can argue
that the zero free energy conditions at GBs (or TJs) must be satisfied
in an average sense. If so, then what will the stable structure look
like if the free energy is positive for some GBs (or TJs) and negative
for others? How will such a structure evolve in time? These seemingly
abstract questions are linked to the practical question of whether
it is worth the efforts to pursue the full stabilization experimentally. 

In our previous publication \citep{Hussein:2024aa}, we developed
a simple model aiming to address some of the questions mentioned above.
The model combines the Ising model on a two-dimensional (2D) lattice
with a lattice gas model of a solid solution. It captures the concurrent
processes of GB migration and solute diffusion, and is solved by kinetic
Monte Carlo (KMC) simulations. The simulations revealed the formation
of a stable polycrystalline (more accurately, bicrystalline) state
in a certain range of temperatures and chemical compositions. The
stable structure is composed of isolated grains nested in one another
and dynamic in nature. The grains are constantly growing and shrinking
and remain in dynamic equilibrium with each other. As expected, the
GB free energy $\gamma$ computed by thermodynamic integration methods
was found to be zero in the equilibrium state. 

Although the previous model \citep{Hussein:2024aa} confirmed the
link between the full stabilization and the $\gamma=0$ condition
and provided some glimpse into the possible equilibrium structures,
it suffered from significant limitations. The model considered only
two crystallographic orientations and could not properly represent
a polycrystalline state. Furthermore, the structures could not contain
TJs, leaving their role in the stabilization unexplored. In addition,
the melting temperature of the single-crystalline state was predicted
to be a second-order transition, not a first-order transition as in
reality.

In this paper, we expand the previous model by replacing the Ising
model with a Potts model \citep{Wu:1982aa} capable of representing
multiple crystallographic orientations. In section \ref{sec:Model-formulation},
we introduce the model and discuss its capabilities. In section \ref{sec:Results},
we apply the model to conduct a parametric study of the alloy phase
diagrams and find the conditions under which a stable polycrystalline
state appears as a domain on a phase diagram. We study the structural
and dynamic properties of the fully stabilized polycrystals and demonstrate
the significant role of TJs in their stabilization. In section \ref{sec:Discussion},
we summarize our findings and put them in perspective with the general
problem of nano-stabilization.

\section{Model formulation\label{sec:Model-formulation}}

We consider a simple square lattice forming a rectangular simulation
block with periodic boundary conditions. The block contains $N$ sites
enumerated by the index $k=1,2,...,N$. Each site is interpreted as
a small crystallite capable of adopting one of $q$ distinct lattice
orientations labeled by index $\sigma=1,2,...,q$. 

For any given distribution of the orientations among the sites, each
site $k$ is characterized by the number $n_{k}$ of nearest-neighbor
sites whose orientations are different from $\sigma_{k}$. Thus,

\begin{equation}
n_{k}=\sum_{(kl)}\left(1-\delta_{\sigma_{k}\sigma_{l}}\right),\label{eq:2}
\end{equation}
where $\delta_{\sigma_{k}\sigma_{l}}$ is the Kronecker delta-function
equal to 1 if $\sigma_{k}=\sigma_{l}$ and 0 otherwise. The symbol
$(kl)$ indicates summation over all neighbors $l$ of site $k$.
If all neighbors have the same orientation as the site $k$, then
$n_{k}=0$. If all neighbors have orientations different from $\sigma_{k}$,
then $n_{k}=4$. 

\begin{figure}[th!]
\centering \includegraphics[width=1\textwidth]{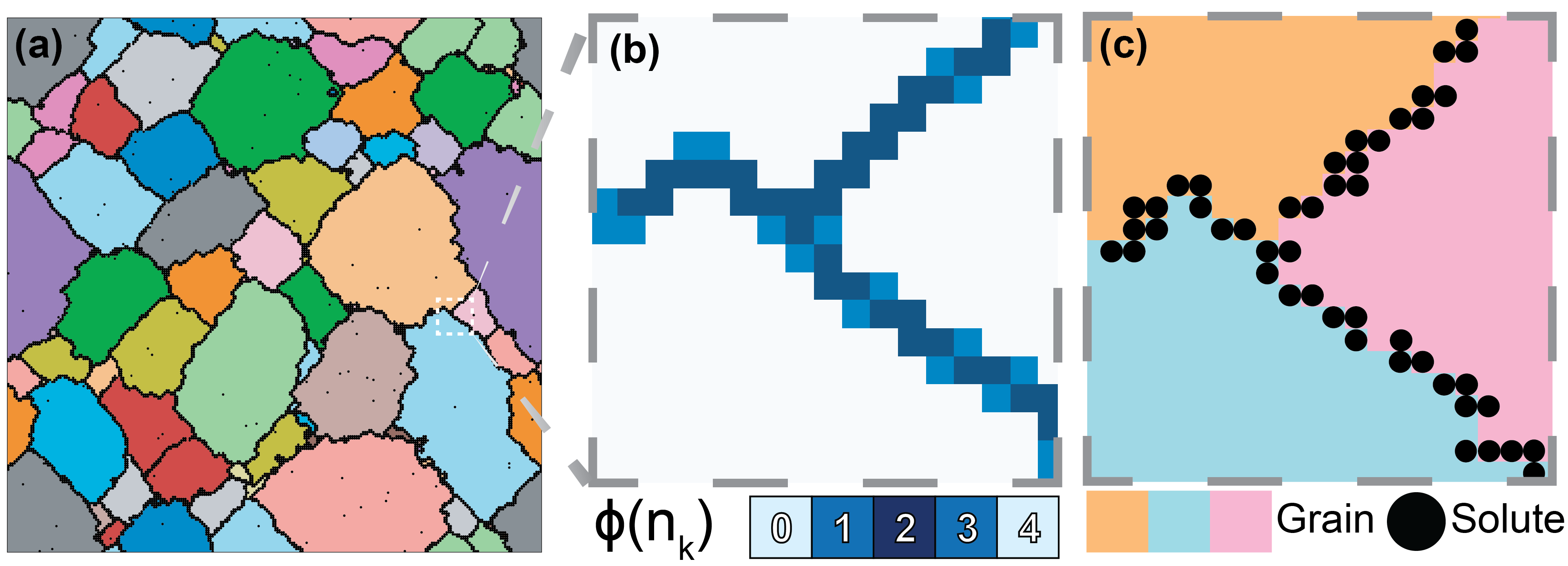} \caption{(a) Snapshot of 256\texttimes 256 polycrystalline system at temperature
$T=0.15$ and solute concentration $c=0.1$. (b) Zoom-in view of a
typical triple junction region with GBs colored according to the function
$\phi(n_{k})$. (c) The same zoom-in region as in (b) showing individual
grains (colored regions) and solute atoms (black circles).}
\label{fig:SystemSetup}
\end{figure}

Under certain conditions discussed later, the system breaks into domains
composed of sites with the same orientation, which we interpret as
grains separated by GBs. An example of a polycrystalline structure
is shown in Fig.~\ref{fig:SystemSetup}(a). In this and other figures
in this paper, the colors represent different crystal orientations
corresponding to different $\sigma$-numbers. Inside the grains, most
sites have $n_{k}=0$ (perfect lattice) and occasionally $n_{k}=4$
(isolated inclusion with a ``wrong'' orientation) or $n_{k}=1$ (neighbors
of the isolated inclusion). In contrast, most sites in GB regions
have $n_{k}=2\pm1$. To identify GBs and track their motion, we introduce
the GB locator function 
\begin{equation}
\phi(n_{k})=1-\frac{1}{4}\left(n_{k}-2\right)^{2},\label{eq:phi}
\end{equation}
which reaches the maximum value of $1$ at sites with $n_{k}=2$.
Fig.~\ref{fig:SystemSetup}(b) depicts a zoom-in portion of a polycrystalline
structure with GBs revealed as sites with large $\phi(n_{k})$ values.
The image illustrates the effectiveness of the automated GB detection
based on the function $\phi(n_{k})$. 

Next, we postulate that pairs of nearest-neighbor sites interact with
a positive energy $J_{gg}>0$ (repulsion) if their orientations are
different and do not interact if the orientations are equal. The repulsive
interactions drive the system into a polycrystalline state with relatively
narrow GBs. The crystallographic energy of the system caused by such
interactions is 
\begin{equation}
E_{\mathrm{cryst}}=\sum_{k}J_{gg}n_{k}\label{eq:1}
\end{equation}
with $E_{\mathrm{cryst}}=0$ in the perfect single-crystalline state.

To describe solute thermodynamics, we use a lattice gas model with
nearest-neighbor interactions. The solute atoms are distributed over
the sites with a single atom per site. Each distribution is characterized
by a set of occupation numbers $\xi_{k}$ with $\xi_{k}=1$ if the
site $k$ is occupied by a solute atom and $\xi_{k}=0$ otherwise.
The average solute concentration in the system is $c=\sum_{k}\xi_{k}/N$.
We further assume that GBs create a potential field attracting the
solute atoms. This field is represented by the following term in the
total energy: 
\begin{equation}
E_{sg}=\sum_{k}\xi_{k}J_{s}+\sum_{k}\xi_{k}J_{sg}\phi(n_{k}),\label{eq:solute-GB-1}
\end{equation}
where $J_{s}$ has the meaning of the solute energy inside the grains.
The product $\xi_{k}\phi(n_{k})$ identifies GB sites occupied by
a solute atom and shifts their energy by the amount $J_{sg}<0$, creating
a driving force for GB segregation. The parameter $J_{sg}$ controls
the strength of GB segregation (Fig.~\ref{fig:SystemSetup}(c)) and
is referred to as the segregation energy. 

To account for solute-solute interactions, we add the following terms
to the total energy:

\begin{equation}
E_{ss}=\frac{1}{2}\sum_{k}\sum_{(kl)}\xi_{k}\xi_{l}J_{ss}+\frac{1}{2}\sum_{k}\sum_{(kl)}\xi_{k}\xi_{l}J_{ssg}\phi(n_{k,l}).\label{eq:s-s}
\end{equation}
Here, the products $\xi_{k}\xi_{l}$ locate pairs of nearest-neighbor
atoms and the factor of $1/2$ eliminates their double-count. If both
atoms are located inside a grain, their interaction energy is $J_{ss}$.
If they are located in a GB region, the interaction energy is additionally
shifted by the amount $J_{ssg}\phi(n_{k,l})$, where the GB locator
$\phi(n_{k,l})$ depends on the average number of misoriented neighbors:
\begin{equation}
n_{k,l}=\frac{n_{k}+n_{l}}{2}.\label{eq:nkl}
\end{equation}
Note that the two terms in Eq.(\ref{eq:s-s}) enable us to control
of the solute-solute interactions inside the grains and in GB segregation
atmospheres separately.

Putting all pieces together, the system energy $E_{i}$ in each microstate
$i=\{\sigma_{k},\xi_{k}\}$ is the sum of three contributions:
\begin{equation}
E_{i}=E_{\mathrm{cryst}}+E_{sg}+E_{ss},\label{eq:E_tot}
\end{equation}
where the crystallographic energy $E_{\mathrm{cryst}}$, the solute-GB
interaction energy $E_{sg}$, and the solute-solute interaction energy
$E_{ss}$ are given by Eqs.~(\ref{eq:1}), (\ref{eq:solute-GB-1}),
and (\ref{eq:s-s}), respectively. The total energy depends on five
physical parameters: $J_{gg}$, $J_{s}$, $J_{sg}$, $J_{ss}$, and
$J_{ssg}$. 

To evolve the system in time, we implement KMC simulations, in which
the system transitions between microstates by overcoming energy barriers
by thermal fluctuations. We adopt the harmonic transition state theory
\citep{Vineyard:1957vo}, in which the transition rate $\nu_{ij}$
from microstate $i$ to microstate $j$ is given by

\begin{equation}
\nu_{ij}=\nu_{0}\exp\left(-\dfrac{\varepsilon_{ij}}{k_{B}T}\right),\label{eq:1.01-1-1}
\end{equation}
where $\varepsilon_{ij}$ is the transition barrier, $k_{B}$ is Boltzmann's
constant, and $\nu_{0}$ is the attempt frequency. The transition
barrier $\varepsilon_{ij}$ depends on the energy difference, $E_{ij}=E_{j}-E_{i}$,
between the states. We adopt the nonlinear energy-barrier relation
\citep{Mishin:2023aa,Mishin:2023ab,Hussein:2024aa}:
\begin{equation}
\varepsilon_{ij}=\begin{cases}
\varepsilon_{0}\exp\left(\dfrac{E_{ij}}{2\varepsilon_{0}}\right), & E_{ij}\leq0,\\
E_{ij}+\varepsilon_{0}\exp\left(-\dfrac{E_{ij}}{2\varepsilon_{0}}\right), & E_{ij}>0
\end{cases}\label{eq:1.02-1}
\end{equation}
where $\varepsilon_{0}$ is the unbiased energy barrier (when $E_{j}=E_{i}$).
This relation ensures that transitions to higher (lower) energy microstates
have a higher (lower) barrier. If the destination microstate has a
much lower energy than the current microstate, the transition barrier
is exponentially small but never becomes zero. In other words, a strong
driving force can suppress the transition barrier but never makes
it exactly zero. 

We consider two types of transitions, which we call flips and jumps.
In a flip transition, a chosen site $k$ switches its current orientation
to one of the remaining $q-1$ orientations with $\sigma_{l}\neq\sigma_{k}$.
If a group of sites flips into the same orientation on one side of
a GB, this can cause GB migration. In a jump event, a chosen solute
atom jumps to a nearest-neighbor site if the latter is unoccupied.
Such jumps represent solute diffusion. The flip and jump transitions
are assigned unbiased barriers $\varepsilon_{0}^{g}$ and $\varepsilon_{0}^{s}$,
respectively. The flip barrier $\varepsilon_{0}^{g}$ is the same
for all sites and is related to the activation energy of GB migration.
The jump barrier $\varepsilon_{0}^{s}$ represents the activation
energy of solute diffusion. In the present model, $\varepsilon_{0}^{s}$
is assigned different values in GBs and inside the grains. Specifically,
the unbiased barrier of a jump is 
\begin{equation}
\varepsilon_{0}^{s}=\varepsilon_{00}^{s}\left[1-\eta\phi(n_{k,l})\right],\label{eq:Solute-barrier-1}
\end{equation}
where $\varepsilon_{00}^{s}$ is the jump barrier inside the grains
and $n_{k,l}$ is the average number of misoriented neighbors in the
initial ($k$) and final ($l$) positions of the solute atom. The
term in square brackets in Eq.(\ref{eq:Solute-barrier-1}) lowers
the barrier for jumps within GB regions. Since in GB environments
$\phi(n_{k,l})\approx1$, the barrier becomes approximately $\varepsilon_{0}^{s}\approx\varepsilon_{00}^{s}\left[1-\eta\right]$.
The parameter $0\leq\eta<1$ controls the decrease in the diffusion
barrier leading to accelerated GB diffusion (``short-circuit'' diffusion
\citep{Kaur95,Chesser:2024aa}) relative to the perfect lattice. In
the simulations, we usually take $\eta=1/2$ to match the experimentally
established correlation $\varepsilon_{0}^{s}\approx\varepsilon_{00}^{s}/2$
between the activation energies of GB and lattice diffusion \citep{Kaur95}.
Note that a solute jump is generally accompanied by a change in the
total energy ($E_{ij}\neq0$). As a result, the actual jump barrier
$\varepsilon_{ij}$ is biased relative to $\varepsilon_{0}^{s}$ according
to Eq.(\ref{eq:1.02-1}).

In the KMC simulations, the number of solute atoms in the system is
kept fixed while changes in the site orientations are unconstrained.
Thus, the statistical ensemble is canonical for the solute atoms and
grand-canonical with zero chemical potential for the site orientations.
A rejection-free ($n$-way \citep{bortz1975new}) KMC algorithm is
implemented. Technical details of this algorithm were presented in
our previous publication \citep{Hussein:2024aa}. In short, at each
step, we apply Eqs.~(\ref{eq:1.01-1-1}) and (\ref{eq:1.02-1}) to
compute the rates of all possible flips at all sites and all possible
jumps of all solute atoms. Normalization of these rates gives us the
transition probabilities to all microstates $j$ accessible from the
given microstate $i$. A random number then selects a new microstate
in proportion to its transition probability. The clock is advanced
according to the total escape rate from the microstate $i$, and the
process repeats from the new microstate. 

Two features distinguish our KMC algorithm from previous Monte Carlo
simulations using the Ising and Potts models with solutes \citep{Mendelev:2001aa,Mendelev:2001wk,Trelewicz2009,Chookajorn2014,Liu:1998ab,Liu:1999aa}.
Firstly, we replace the Metropolis scheme with the probability calculations
using the actual transition barriers. This scheme is more suitable
for the modeling of solute diffusion and GB migration than the Metropolis
algorithm. The latter correctly samples canonical fluctuations in
equilibrium states but the path toward equilibrium is not guaranteed
to be physically meaningful. Secondly, previous simulations within
the Potts model only allowed flips to one of the neighboring orientations.
In contrast, we allow the current orientation to switch to \emph{any}
of the alternative $q-1$ orientations. This imposes fewer constraints
on the system evolution and captures processes such as new grain nucleation
and orientational fluctuations in single crystals as well as disordered,
liquid-like structures.

The KMC simulations were performed in normalized variables whose relations
to the physical variables are summarized the Appendix. All energies
are normalized by $J_{gg}$, and the unit of time is $\nu_{0}^{-1}$.
The attempt frequency $\nu_{0}$ is assumed to be the same for all
transitions. From now on, we only use the normalized variables denoted
by the same symbols as the respective physical variables. In most
simulations reposted below, we assume $J_{s}=0$.

\section{Results\label{sec:Results}}

\subsection{The melting transition in the solute-free system}

\begin{figure}
\begin{centering}
\includegraphics[totalheight=0.87\textheight]{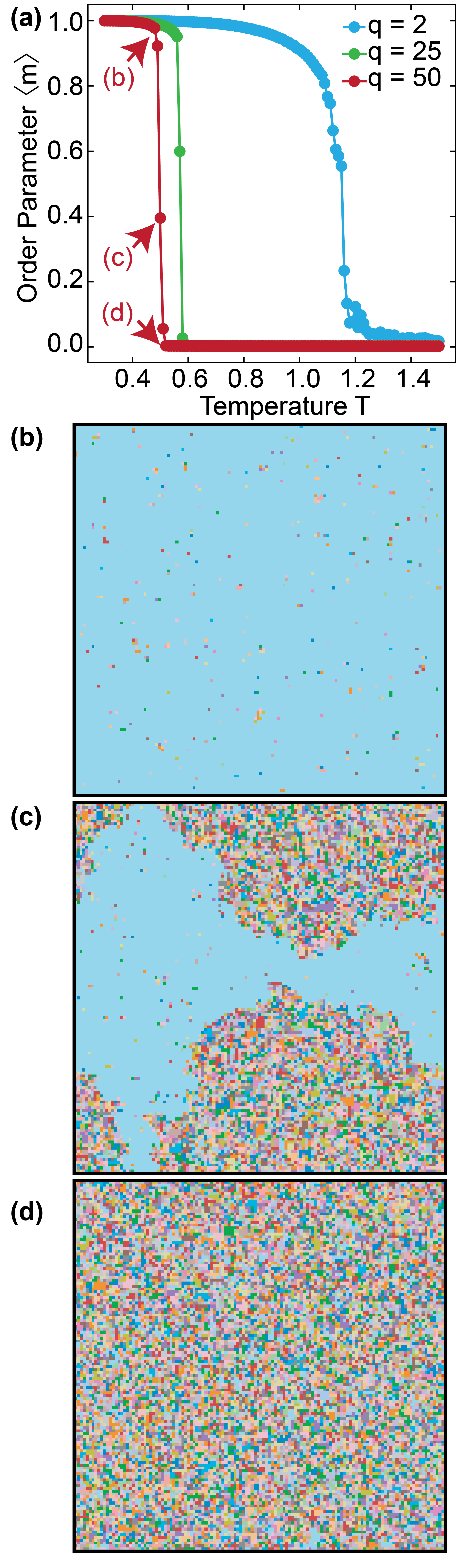}
\par\end{centering}
\caption{(a) Phase diagram of solute-free Potts model computed on a $128\times128$
grid for three representative $q$-values. (b) Single crystalline
state with occasional isolated inclusions below the critical point.
(c) Transient two-phase state showing solid-liquid interphase boundaries.
(d) Fully disordered (liquid-like) state above the critical point.\label{fig:Pott-solute-free}}
\end{figure}

Before studying the alloy systems, we performed solute-free simulations
to validate our methodology and software. Fig.~\ref{fig:Pott-solute-free}(a)
shows the computed orientational order parameter $\left\langle m\right\rangle $
as a function of temperature for three different $q$ values. The
order parameter was computed by \citep{Binder:1981aa} 
\begin{equation}
\left\langle m\right\rangle =\left\langle \dfrac{(N_{m}/N)q-1}{q-1}\right\rangle ,\label{eq:m}
\end{equation}
where $N_{m}$ is the number of sites with the dominant orientation
and the angular brackets represent averaging over many KMC steps.
Below a critical point, a system initiated with a single orientation
remains predominantly in this orientation ($\left\langle m\right\rangle \approx1$)
except for a small concentration of misoriented sites and their small
clusters (Fig.~\ref{fig:Pott-solute-free}(b)). The local orientation
flips are caused by thermal fluctuations and are revealed by our simulations
because we allow flip transitions to all orientations without restricting
them to nearest neighbors. Above the critical temperature $T_{c}$,
the system becomes fully disordered ($\left\langle m\right\rangle \rightarrow0$)
with weak short-range dynamic correlations among the site orientations
(Fig.~\ref{fig:Pott-solute-free}(d)). 

In the magnetic terminology, the states below and above $T_{c}$ are
ferromagnetic and paramagnetic, respectively. Since we interpret the
indices $\sigma_{k}$ as crystallographic orientations, we refer to
the two states as a single crystal and liquid, respectively. Accordingly,
the critical temperature is interpreted as the melting point, which
in the 2D Potts model with $q>4$ is a first-order phase transition
\citep{Chen:2023aa,Beffara:2012aa}. Since the ensemble is grand canonical,
only single-phase states are possible as equilibrium states. However,
the simulations also capture transient two-phase states as the system
evolves from one phase to the other. An example of a transient solid-liquid
state is shown in Fig.~\ref{fig:Pott-solute-free}(c). The relatively
sharp interphase boundaries are consistent with the first-order nature
of melting in this model.

The computed values of $T_{c}$ closely follows the analytical solution
of the Potts model for the simple square lattice \citep{Chen:2023aa,Beffara:2012aa}:
\[
T_{c}=\dfrac{1}{\ln(1+\sqrt{q})}.
\]
As the number of orientations $q$ increases, the melting temperature
decreases (Fig.~\ref{fig:Pott-solute-free}(a)), which is consistent
with the increased configurational entropy of the liquid phase. Note
that the discontinuity of the order parameter at the melting point
sharply increases with $q$.

\subsection{The alloy phase diagrams}

In the presence of solute atoms, the phase diagrams become more complex.
A parametric study was performed to explore representative regions
of the large parameter space of the model. The results were presented
in the form of temperature-composition phase diagrams showing the
domains of single phases and phase coexistences. For each set of model
parameters, a large number of simulations were performed on a grid
of temperatures and solute concentrations between $c=0$ to $c=1$.
For each temperature-composition pair, a $64\times64$ system was
equilibrated by KMC simulations until all properties reached time-independent
values. Large solute diffusivities were implemented to accelerate
the convergence. Different initial states were tested to verify that
the system always arrives at the same final state. Several structural
and dynamic properties were computed to identify the phases automatically
in addition to visual inspection. Since this construction has a discrete
character, the lines delineating different domains on the phase diagrams
are only accurate to within the temperature-composition grid size
$\Delta T\times\Delta c$, which was typically $0.02\times0.1$. Supplementary
Figure 1 presents a typical phase diagram together with the grid used
for its construction. Supplementary Figure 2 shows typical structures
of all phases appearing on the phase diagrams alongside the convergence
plots demonstrating that these structures do not depend on the initial
state of the KMC simulations.

We first discuss the simplest case when the solute atoms are attracted
to GBs ($J_{sg}=-1.6$) to form segregation atmospheres but do not
interact with each other ($J_{ss}=J_{ssg}=0$). Formally, the solute
atoms form a perfect solution. In reality, they still interact with
each other indirectly through coupling to the orientational order
parameter. Indeed, although the solute atoms are primarily attracted
to GBs, which are identified using the GB locator function $\phi(n_{k})$,
isolated misoriented sites and their small clusters can also act as
attractors because they have nonzero $\phi(n_{k})$ values. Occasionally,
the solute atoms confuse such misoriented inclusions with GBs and
are attracted to them. For example, if a site inside a single-crystalline
region flips into a ``wrong'' orientation, it can attract two solute
atoms, which will then try to remain neighbors of that site. The imposed
positional correlations between pairs of solute atoms acts as an effective
attractive interaction. In most cases, such interactions are overshadowed
by the strong attraction to GBs. However, they can still affect the
shape of the phase diagram. There is also a reciprocal effect, in
which the solute atoms promote orientational disorder among their
neighbors due to their attraction to misoriented sites. In particular,
the solute atoms are attracted to liquid environments and stabilize
the liquid phase.

The phase diagram in Fig.~\ref{fig:Phase-diagram-1} features a single-crystalline
phase, a liquid solution phase, and a solid-liquid coexistence domain
between the solidus and liquidus lines. The negative slope of the
liquidus line is consistent with the solute-induced liquid stabilization
mentioned above. The images next to the phase diagram display typical
alloy structures with color-coded crystallographic orientations and
the solute atom distributions shown by black dots. The bottom part
of the diagram represents kinetically trapped states that could not
be fully equilibrated on the time scale of our simulations. 

According to the diagram in Fig.~\ref{fig:Phase-diagram-1}, the
ground state of the solid solution is a single crystal. A polycrystalline
state is unstable and must transform into a single crystal. To verify
this prediction, we chose a classical polycrystalline structure as
the initial state of KMC simulations (Fig.~\ref{fig:Circular-grain}(a)).
The temperature and composition were selected within the single-crystalline
domain on the phase diagram. During the simulation, the structure
was found to coarsen by first normal and later abnormal grain growth
(Fig.~\ref{fig:Circular-grain}(b)), and eventually transformed into
a single crystal (Fig.~\ref{fig:Circular-grain}(c)). This behavior
is typical for unstable polycrystals with $\gamma>0$.

We find that the liquid structure undergoes marked changes with decreasing
temperature and increasing solute concentration. At high temperatures
and low-to-moderate solute concentrations, the liquid solution is
similar to the solute-free liquid (compare Figs.~\ref{fig:Pott-solute-free}(d)
and \ref{fig:Phase-diagram-1}). The orientations $\sigma_{k}$ randomly
fluctuate from one site to the next with little correlation. Sites
with $n_{k}\geq3$ dominate. When more solute is added and/or the
temperature is decreased, the liquid develops a more coarse-grained
structure with stronger spatial correlations between neighboring orientations.
The sites with $n_{k}\leq2$ become prevalent. This structure can
be described as a collection of orientational domains a few inter-site
spacings in size, which constantly appear and disappear. We call this
form of liquid a ``supercooled liquid''. Several structural characteristics
were computed to quantify the difference between the two liquids.
For example, Supplementary Figure 3 shows a set of histograms of $n_{k}$
values in the $c=1$ alloy. The histograms were obtained by averaging
over long KMC runs after the system reached equilibrium. As temperature
decreases, the histograms shift towards smaller $n_{k}$ values and
reverse their skewness from right to left. These trends are quantified
in Fig.~\ref{fig:liquid}(a,b), showing the average number of unlike
neighbors $\overline{n}_{k}$ and the skewness parameter as functions
of temperature. The plots show that $\overline{n}_{k}$ changes its
behavior and the skewness reverses its sign at $T\approx0.35$. 

In addition to the structural distinctions, the supercooled liquid
displays much slower dynamics compared to the high-temperature liquid.
For example, Fig.~\ref{fig:liquid}(c) shows the Arrhenius plot of
the orientation switching rate $r$ as a function of temperature at
a fixed chemical composition. To calculate $r$, we keep track of
the orientation changes at all sites during an equilibrium KMC run.
At each site, we compute the time interval $\Delta t$ between the
moment when the site switched to its current orientation and the moment
when it switches to a new orientation. After the switch, we reset
the clock and start counting the time to the next switch, and so on.
The value $\overline{\Delta t}$ averaged over a long KMC run and
all sites represents the time scale on which the sites preserve their
orientation. Accordingly, $r=1/\overline{\Delta t}$ characterizes
the rate of orientational changes in the liquid, which is a measure
of its intrinsic dynamics. The plot of $\log r$ versus $1/T$ shown
in Fig.~\ref{fig:liquid}(c) displays a non-Arrhenius behavior. The
increase in $r$ with temperature accelerates above $T\approx0.35$,
which is approximately where the structural changes in the liquid
are observed. We associate the structural and dynamical changes in
the liquid at $T\approx0.35$ with the formation of supercooled liquid.
The temperature of this transition depends on the chemical composition.
Because this transition is continuous, its position on the phase diagram
is approximate and is shown by a dashed line. This line separates
the stable polycrystalline state from a single crystal with inclusions
assuming a threshold grain size of 12 sites.

Next, we keep the same segregation energy ($J_{sg}=-1.6$) but turn
on the solute-solute interactions making them attractive. For example,
we set $J_{ss}=-0.25$ in the grains and $J_{ssg}=-0.05$ at GBs.
The phase diagram (Fig.~\ref{fig:Phase-diagram-2}) is qualitatively
similar to the previous one, except that the solid solubility is markedly
smaller and the liquidus line is shifted towards higher temperatures,
opening a wider miscibility gap. It becomes especially clear that
the supercooling transition in the liquid changes the solid-liquid
interphase boundaries in two-phase alloys. While at high temperatures
such boundaries are rather fluffy, below the supercooling transition
they become remarkably sharp. 

The most interesting case arises when the segregation energy remains
the same as above ($J_{sg}=-1.6$) but the solute-solute interactions
become repulsive (e.g., $J_{ss}=0.25$ in the grains and $J_{ssg}=0.05$
at GBs). The phase diagram changes substantially (Fig.~\ref{fig:Phase-diagram-3}).
The solidus line shifts towards larger solute concentrations, narrowing
the miscibility gap. Most importantly, two different solid-solution
structures can now exist under the solidus line: the previously observed
single-crystalline phase and a new stable polycrystalline phase. The
transition between the single-crystalline and polycrystalline structures
is continuous and is marked on the phase diagram with a dashed line. 

\subsection{The stable polycrystalline state}

In this section, we examine the stable polycrystalline structure in
more detail. The equilibrium nature of this structure was confirmed
by restarting KMC simulations from different initial states and verifying
that we always arrive at the same polycrystalline structure. One of
the convergence tests is demonstrated in Fig.~\ref{fig:Structure-evolution}.
KMC simulations were performed at the temperature of $T=0.1$ and
the chemical composition of $c=0.2$. This state falls into the polycrystalline
field on the phase diagram (Fig.~\ref{fig:Phase-diagram-3}). In
simulations starting from a single crystal with a random solute distribution,
the solute atoms create a set of small misoriented clusters, which
then grow and eventually form a set of isolated grains embedded in
the matrix of the initial grain. In the final state, most of the solute
atoms are located at the boundaries of the embedded grains. The grain
size fluctuates around a time-independent average value. Alternatively,
simulations were started from a random distribution of the solute
atoms and orientations over the sites. This initial structure is similar
to the high-temperature liquid solution. First, randomly oriented
crystalline grains nucleate at multiple locations, grow, and eventually
impinge on each other to form a classical polycrystalline structure
with strong solute segregation. This structure is unstable and coarsens
with time following the classical grain-growth scenario. At some point,
the grain growth becomes abnormal, with one grain starting to consume
all other grains. However, in contrast to the classical behavior,
a set of isolated grains remains embedded in the growing grain, until
the latter spreads over the entire simulation box and becomes a matrix
grain. The final state of this evolution is the same as when starting
from the single-crystalline state. The Supplementary Figure 2 shows
that the system energy converges to the same value in both cases.
These and similar tests confirm that the polycrystalline structure
found here is indeed the thermodynamic ground state of the alloy.

We emphasize that the equilibrium polycrystalline structure is not
static. The grains are not frozen in a special configuration that
minimizes the free energy. Instead, this is a state of \emph{dynamic
equilibrium}, in which some grains grow while other grains shrink.
For instance, Fig.~\ref{fig:dynamics}(a) displays a set of snapshots
of an equilibrated polycrystal taken at different moments of time.
While all structures look similar, the positions, sizes, and shapes
of individual grains change from one snapshot to another. It is only
the \emph{average} structural properties of the polycrystal that remain
constant. Figs.~\ref{fig:dynamics}(c,d) demonstrate that, once equilibrium
is reached, all structural and dynamic characteristics of the polycrystal
fluctuate around constant values. Such characteristics include the
average number of atoms in a grain (Fig.~\ref{fig:dynamics}(c))
and the total number of grains in the system (Fig.~\ref{fig:dynamics}(d)),
which display significant variations around constant average values.
In Fig.~\ref{fig:dynamics}(b), we plot the sizes of ten selected
grains as functions of time. The plot tells us a story of the birth,
life and death of the grains. Some new grains appear (e.g., by splitting
from a larger grain or by homogeneous nucleation), grow in size, then
shrink, and some eventually disappear. Some of the nearly extinct
grains start growing again. 

Another prominent feature of the stable polycrystals is the absence
of TJs. To eliminate them, the grains form isolated inclusions within
a matrix grain. Occasionally, nested structures are observed, with
a smaller grain enclosed within a larger. As the grains grow and shrink,
they always stay clear of each other and avoid even touching. This
TJ avoidance suggests a positive TJ free energy. Fig.~\ref{fig:2-grains}(a)
illustrates the TJ avoidance effect. In an alloy known to be a stable
polycrystal in thermodynamic equilibrium, two grains separated by
a planar GB have been artificially embedded in a matrix grain. This
structure contains two triple junctions. During the KMC simulations,
the grains develop extensive protrusions, which later split off as
smaller grains. Eventually, the initial grains fully separate from
each other, eliminating the TJs. When the same initial state is created
in an alloy whose ground state is a single crystal, the scenario is
different (Fig.~\ref{fig:2-grains}(b)). Both grains shrink and eventually
disappear while maintaining contact and thus the two TJs.

\section{Discussion and conclusions\label{sec:Discussion}}

We proposed a simple 2D model of a polycrystalline alloy combining
the Potts model for grain orientations with a lattice-gas model for
solute thermodynamics and diffusion. The model couples the evolution
of the crystal orientations (and thus GB migration) to thermodynamic
properties of the alloy and the solute diffusion. It recognizes that
the thermodynamic parameters and the solute diffusion coefficients
are generally different inside the grains and at GBs. The model goes
beyond the uniform boundary approximation, automatically includes
nucleation processes due to its stochastic nature, and captures the
diffusive time scales not accessible by material-specific methods
such as molecular dynamics simulations. The model is solved by KMC
simulation with nonlinear transition barriers. We have applied this
model to investigate the problem of polycrystalline stability, but
it has a much broader applicability. It can be used to study other
processes, such as solute drag and the impact of solute segregation
on GB diffusion. 

The main outcome of this work is the confirmation that a polycrystalline
alloy can be thermodynamically stable. The initial prediction of full
stabilization \citep{Weissmuller1994,Weissmuller:1993aa} and the
subsequent thermodynamic analyses \citep{Kirchheim:2002aa,Hussein:2024aa,Liu:2004aa,Kirchheim2007a,Kirchheim2007b,Krill:2005aa,Schvindlerman2006}
were based on strong assumptions. For example, they disregarded the
role of TJs and assumed that all GBs were equal. The present model
is free from those assumptions and nevertheless confirms that the
fully stabilized polycrystalline state exist and is consistent with
thermodynamics. 

The KMC simulations within this model helped us realize that the minimum
of the total free energy cannot be achieved by designing a particular
polycrystalline structure with special grain shapes satisfying the
equilibrium conditions among all capillarity vectors at all TJs. Such
a structure is unlikely to exist. Instead, the structure minimizing
the total free energy is in a state of dynamic equilibrium between
the competing processes of grain growth and grain refinement. 

Thermodynamic analyses within the uniform boundary model \citep{Weissmuller:1993aa,Weissmuller1994,Kirchheim:2002aa,Liu:2004aa,Krill:2005aa,Schvindlerman2006,Kirchheim2007a,Kirchheim2007b,Hussein:2024aa}
predict that the total free energy minimum requires that the GB free
energy be zero. The real GBs are not equal, and the $\gamma=0$ condition
is unlikely to be achieved in all of them simultaneously. Our simulations
suggest that the $\gamma=0$ condition should be met in an average
sense, with GBs alternating between positive and negative $\gamma$
values and the polycrystalline structure ``breathing'' by constantly
switching between the grain coarsening and grain refinement.

TJs play a prominent role in the stabilization problem. In thermodynamically
unstable polycrystals, their total free energy is smaller than the
total GB free energy, making the latter the leading driving force
for the grain coarsening. However, when the GB free energy tends to
zero, or simply becomes small enough, the elimination of TJs becomes
the leading driver of the structural evolution. For full stabilization,
their free energy (per unit length) might also be suppressed to zero
by an appropriate solute segregation. However, it seems highly improbable
that this can be achieved at all TJs simultaneously with the $\gamma=0$
condition at all GBs. Most likely, the structure will try to eliminate
the TJ, and the quadruple points with them. This can be achieved by
creating a set of embedded grains within a large matrix grain, which
is exactly the type of structure we see in the simulations.

An important aspect not discussed in this paper is the role solute
diffusion. The latter must be fast enough to ensure that the solute
atoms can catch up with the moving GBs and keep their free energy
close to zero. The role of solute diffusion was preliminary explored
in our previous publication \citep{Hussein:2024aa} but should be
investigated more systematically in the future. 

Although the proposed model does not represent any particular material
and is intended for a generic analysis of thermodynamics and kinetics
of polycrystalline alloys, it was important to map the domain of the
model parameters explored in our simulations onto thermodynamic properties
of real alloy systems. In the Supplementary Information file accompanying
this article, we have expressed the model parameters through the experimentally
accessible properties such as the heat of solution in the dilute limit,
$H_{\mathrm{mix}}$, the GB segregation energy $E_{s}$, the GB free
energy in the pure solvent material $\gamma$, and the regular solution
parameter $\Omega^{\prime}$ inside the GBs (which is generally different
from $\Omega$ within the grains). The estimates are based on two
reasonable assumptions: the area per atom in the GB core is on the
order of 1 nm$^{2}$, and $\gamma$ is about 1 J/m$^{2}$. 

\begin{figure}[h]
\begin{centering}
\includegraphics[width=0.95\textwidth]{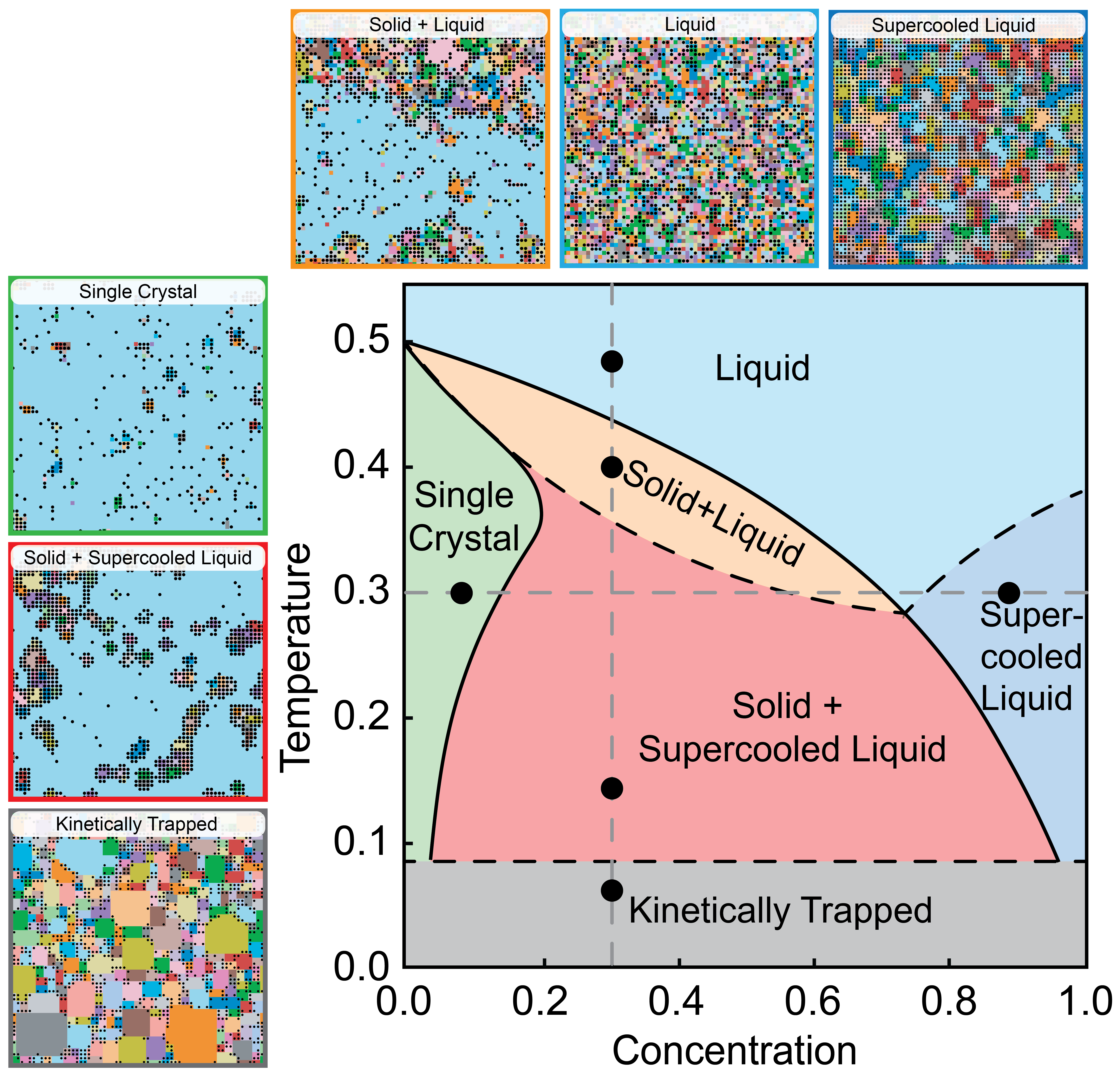}
\par\end{centering}
\caption{Phase diagram when the solute atoms are attracted to GBs ($J_{sg}=-1.6$)
but do not interact with each other ($J_{ss}=J_{ssg}=0$). The images
show the structures corresponding to the black circles on the diagram.
The site orientations are color-coded and the solute atoms are represented
by black dots.\label{fig:Phase-diagram-1}}
\end{figure}

\begin{figure}
\begin{centering}
\includegraphics[width=0.4\textwidth]{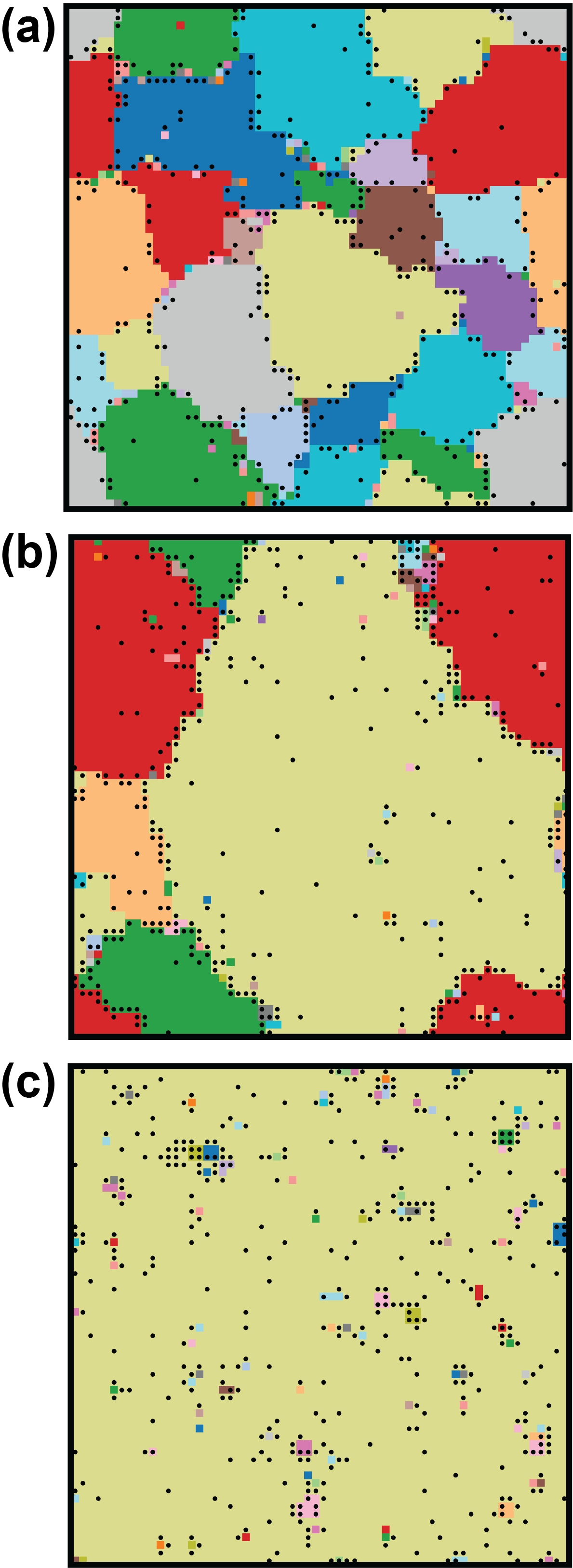}
\par\end{centering}
\caption{Snapshots of KMC simulations of an isolated grain within the single-crystalline
domain on the phase diagram in Fig.~\ref{fig:Phase-diagram-1} ($c=0.1$,
$T=0.36$). (a) Initial state. (b) Intermediate state showing grain
coarsening and beginning of abnormal grain growth. (c) Final single-crystalline
state.\label{fig:Circular-grain}}

\end{figure}

\begin{figure}
\begin{centering}
\includegraphics[totalheight=0.82\textheight]{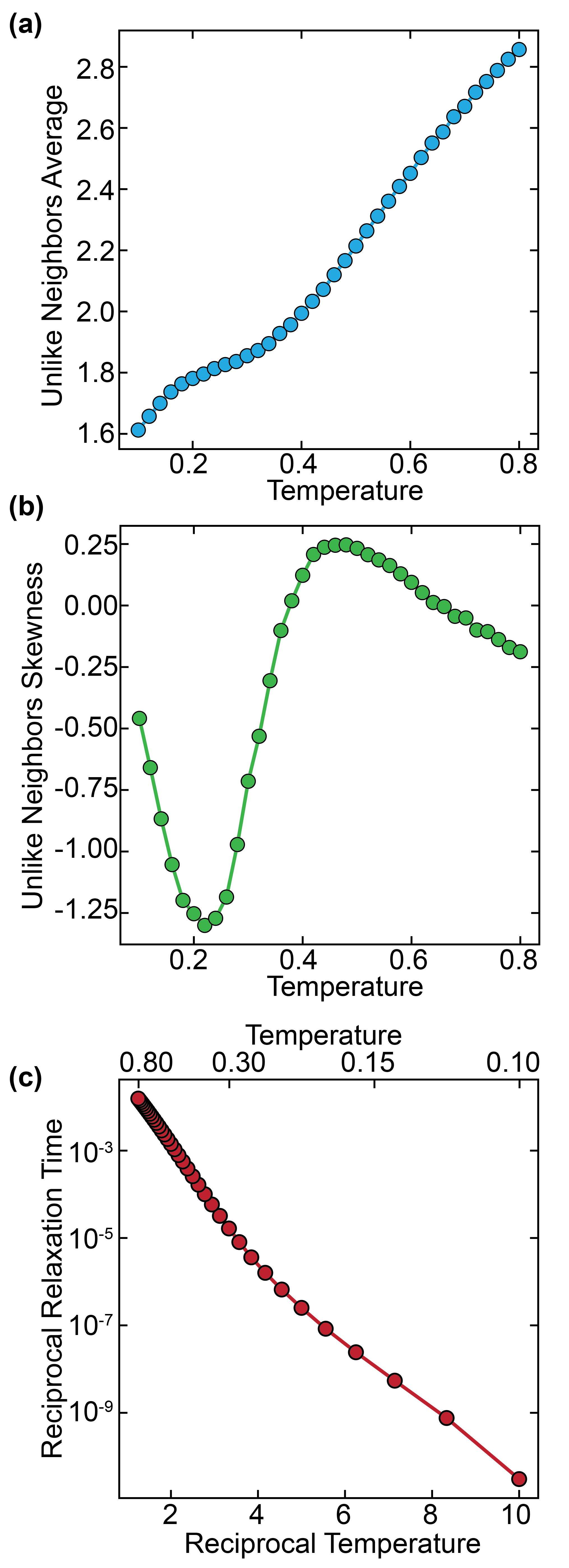}
\par\end{centering}
\caption{Structural and kinetic characteristics of the pure solvent liquid
($c=1$) as a function of temperature. The model parameters are $J_{sg}=-1.6$,
$J_{ss}=J_{ssg}=0$. (a) The average number of unlike neighbors $\overline{n}_{k}$.
(b) The skewness of $n_{k}$ histograms. (c) Arrhenius diagram of
the orientation switching rate $r$.\label{fig:liquid}}

\end{figure}

\begin{figure}[h]
\begin{centering}
\includegraphics[width=0.95\textwidth]{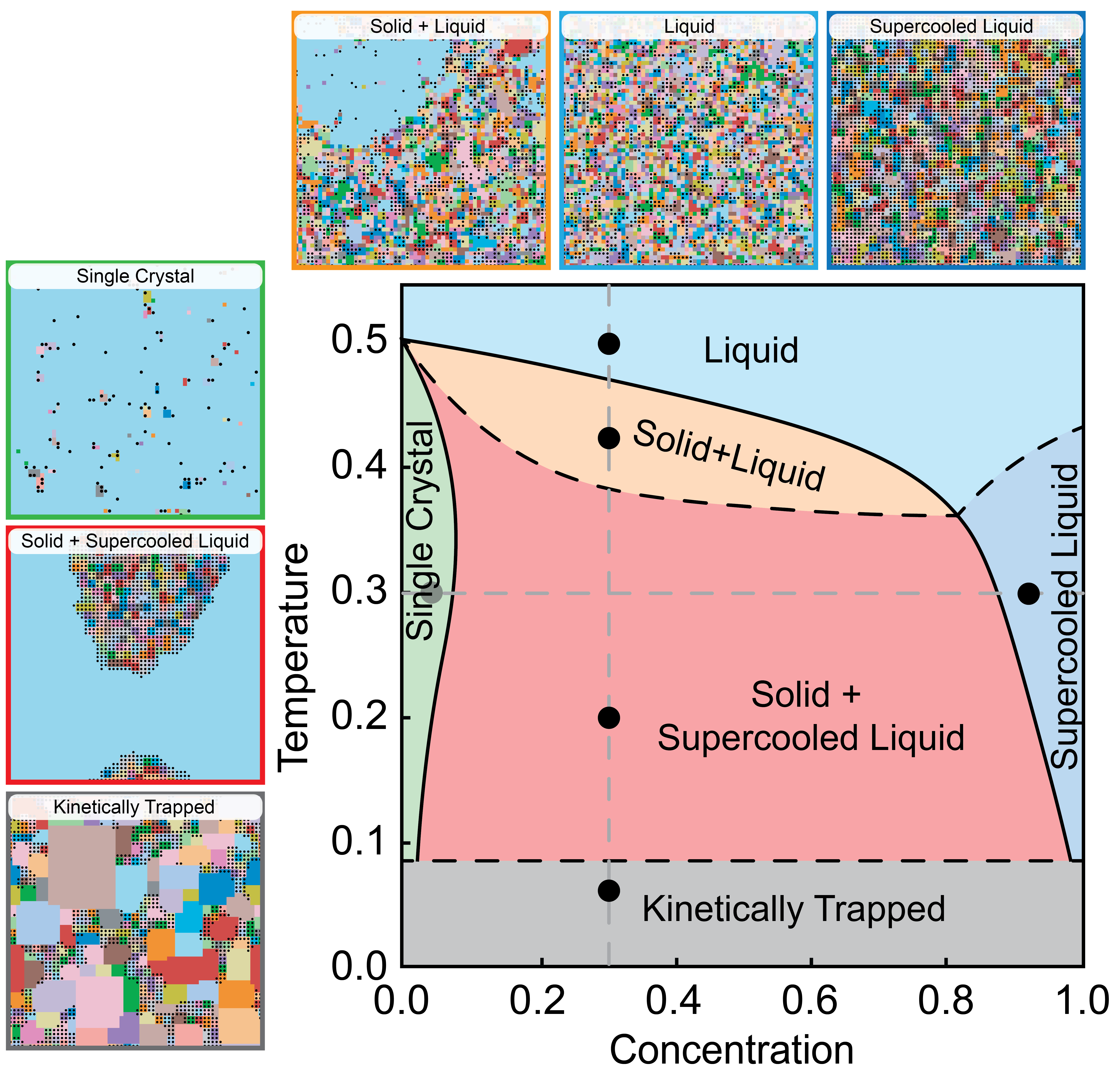}
\par\end{centering}
\caption{Phase diagram when the solute atoms are attracted to GBs ($J_{sg}=-1.6$)
and interact by attractive forces ($J_{ss}=-0.25$, $J_{ssg}=-0.05$
). The images show the structures corresponding to the black circles
on the diagram. The site orientations are color-coded and the solute
atoms are represented by black dots.\label{fig:Phase-diagram-2}}
\end{figure}

\begin{figure}[h]
\begin{centering}
\includegraphics[width=0.95\textwidth]{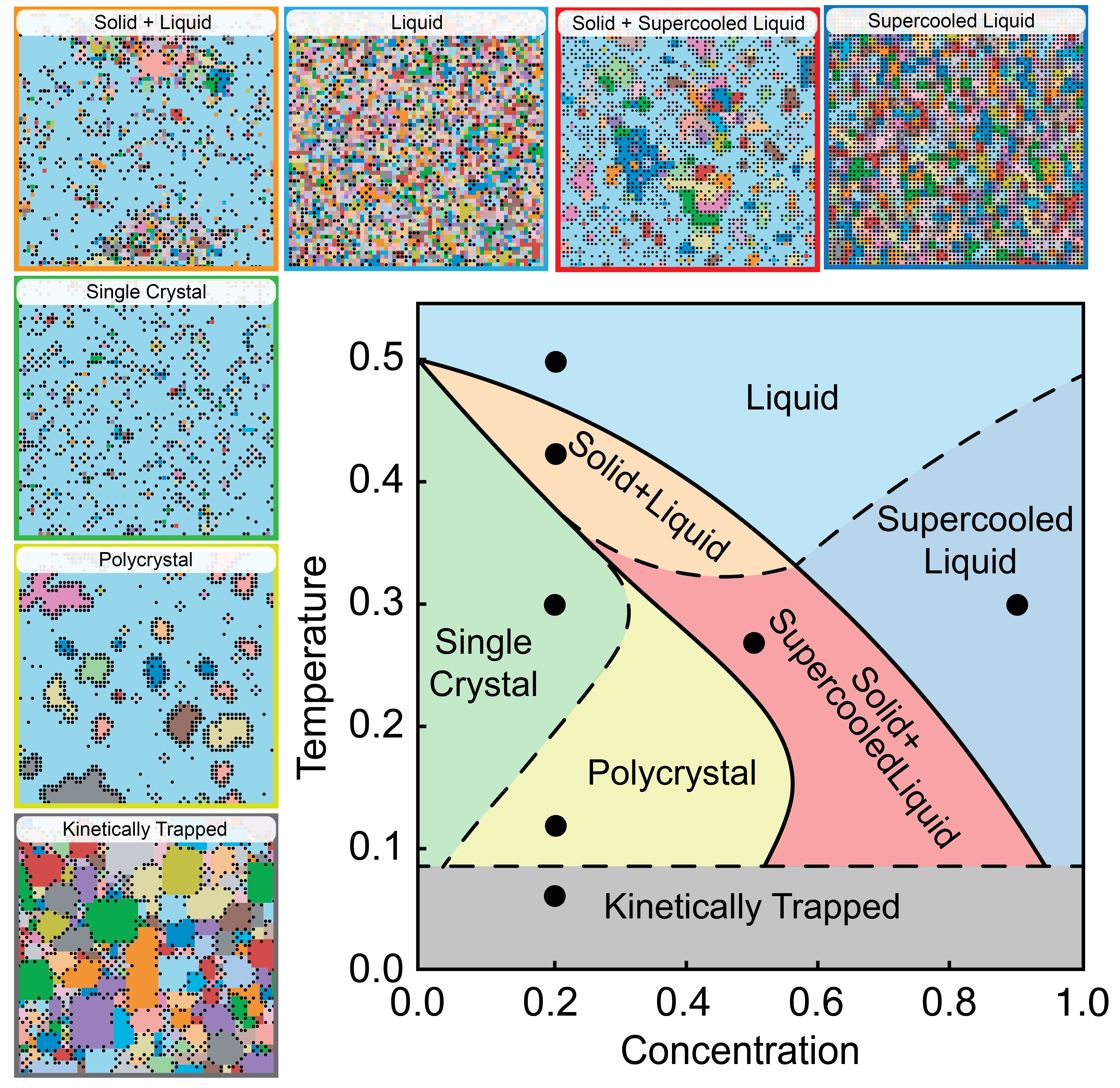}
\par\end{centering}
\caption{Phase diagram when the solute atoms are attracted to GBs ($J_{sg}=-1.6$)
and interact by repulsive forces ($J_{ss}=0.25$, $J_{ssg}=0.05$).
The images show the structures corresponding to the black circles
on the diagram. The site orientations are color-coded and the solute
atoms are represented by black dots.\label{fig:Phase-diagram-3}}
\end{figure}

\begin{figure}
\includegraphics[width=1\textwidth]{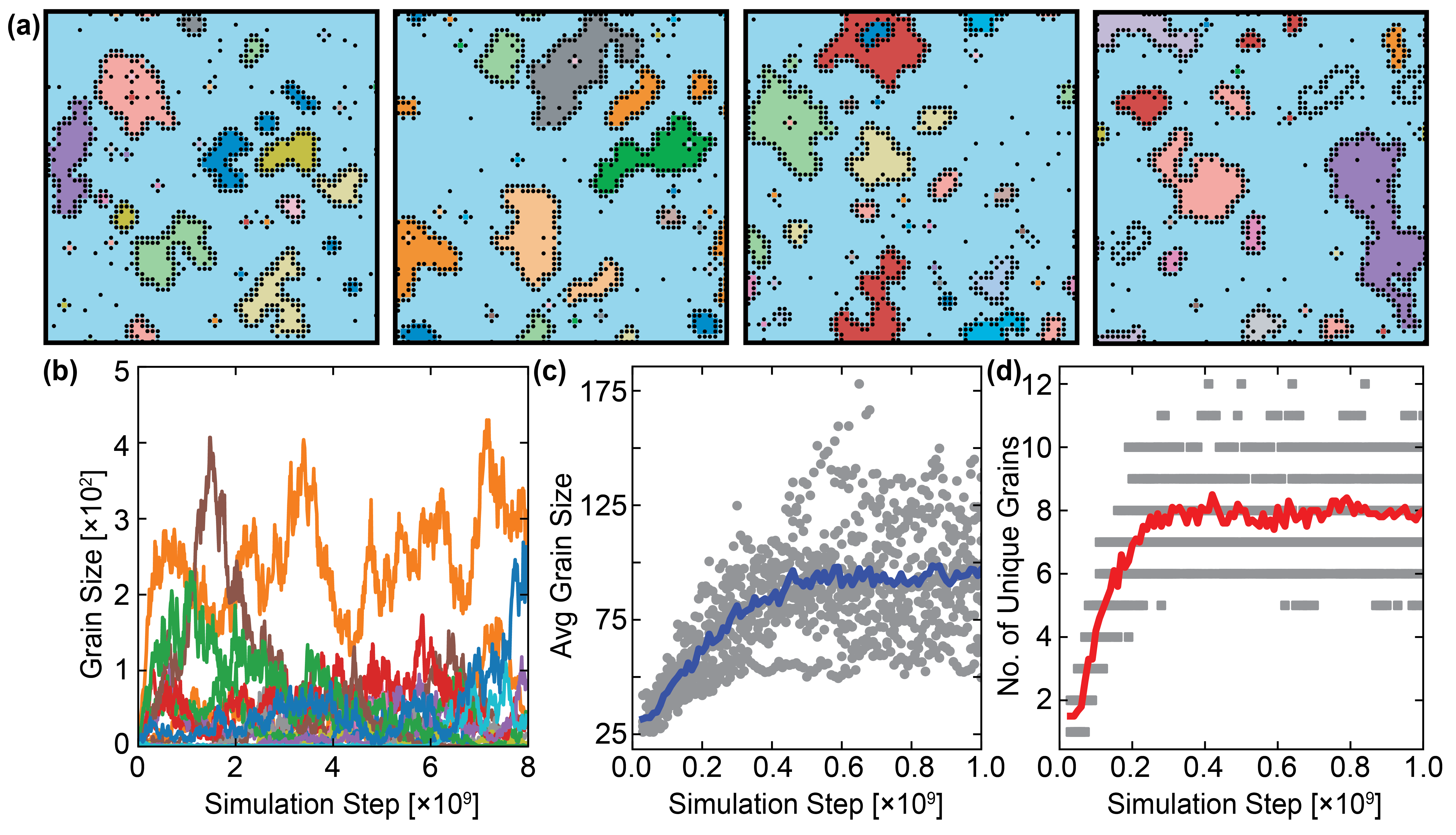}

\caption{Structure evolution (left to right) of the alloy with the composition
$c=0.2$ at temperature $T=0.1$ starting from (a) single crystal
and (b) randomly distributed site orientations. The model parameters
are the same as for the phase diagram in Fig.~\ref{fig:Phase-diagram-3}.
In both cases, the system convergences to the stable polycrystal.\label{fig:Structure-evolution}}
\end{figure}

\begin{figure}
\includegraphics[width=1\textwidth]{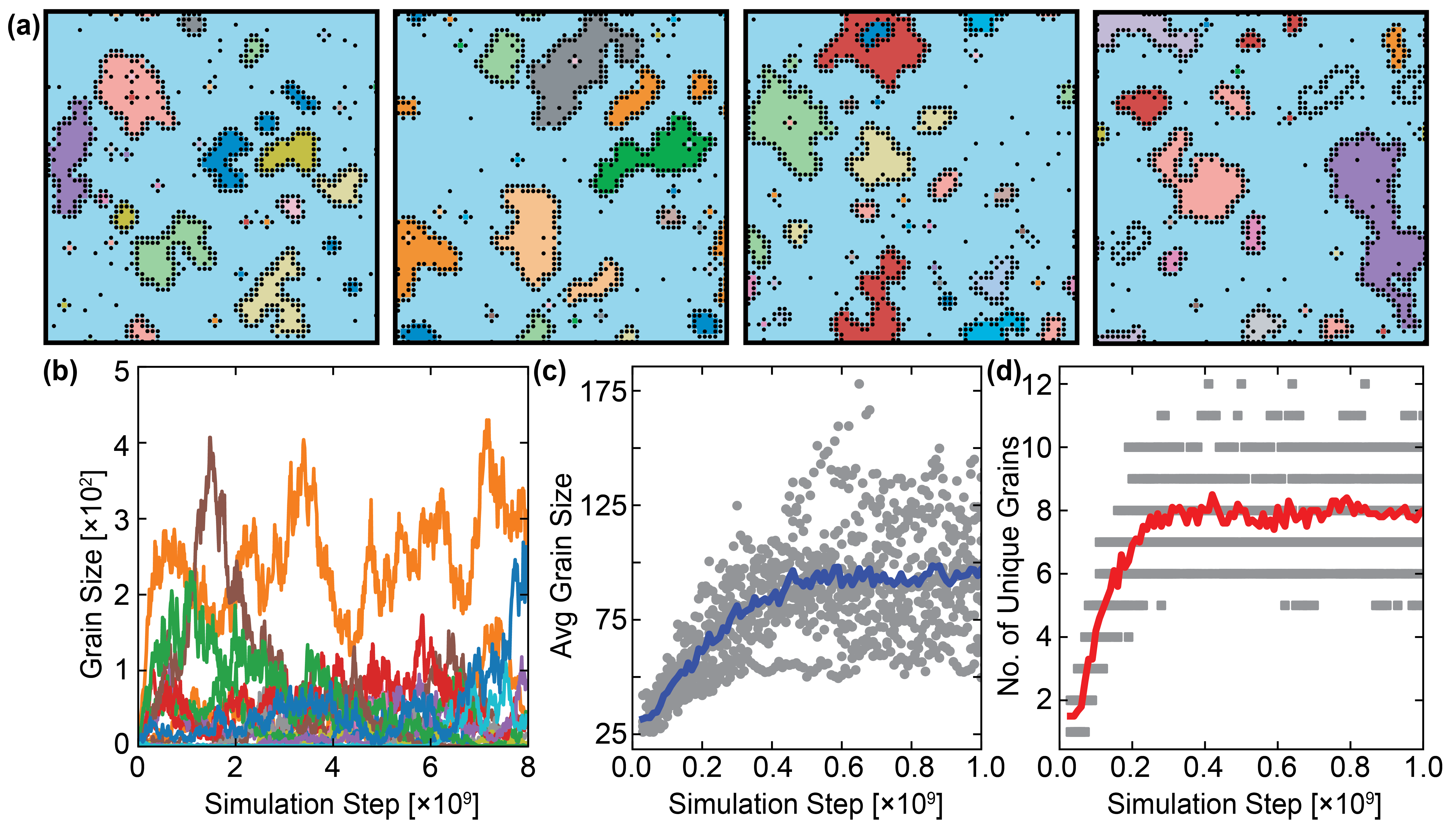}

\caption{(a) Snapshots of equilibrium configurations during KMC simulations
of a stable polycrystalline alloy. (b) Grain size evolutions for ten
selected grains, showing their shrinkage and growth. (c) The sizes
of individual grains in ten KMC runs as a function of time. The blue
curve represents the average grain size. (d) The number of unique
grains as a function of time in ten KMC runs. The red curve represents
the average number of grains. The model parameters are $J_{sg}=-1.6$,
$J_{ss}=0.25$, $J_{ssg}=0.05$.\label{fig:dynamics}}

\end{figure}

\begin{figure}
\begin{centering}
\includegraphics[width=1\textwidth]{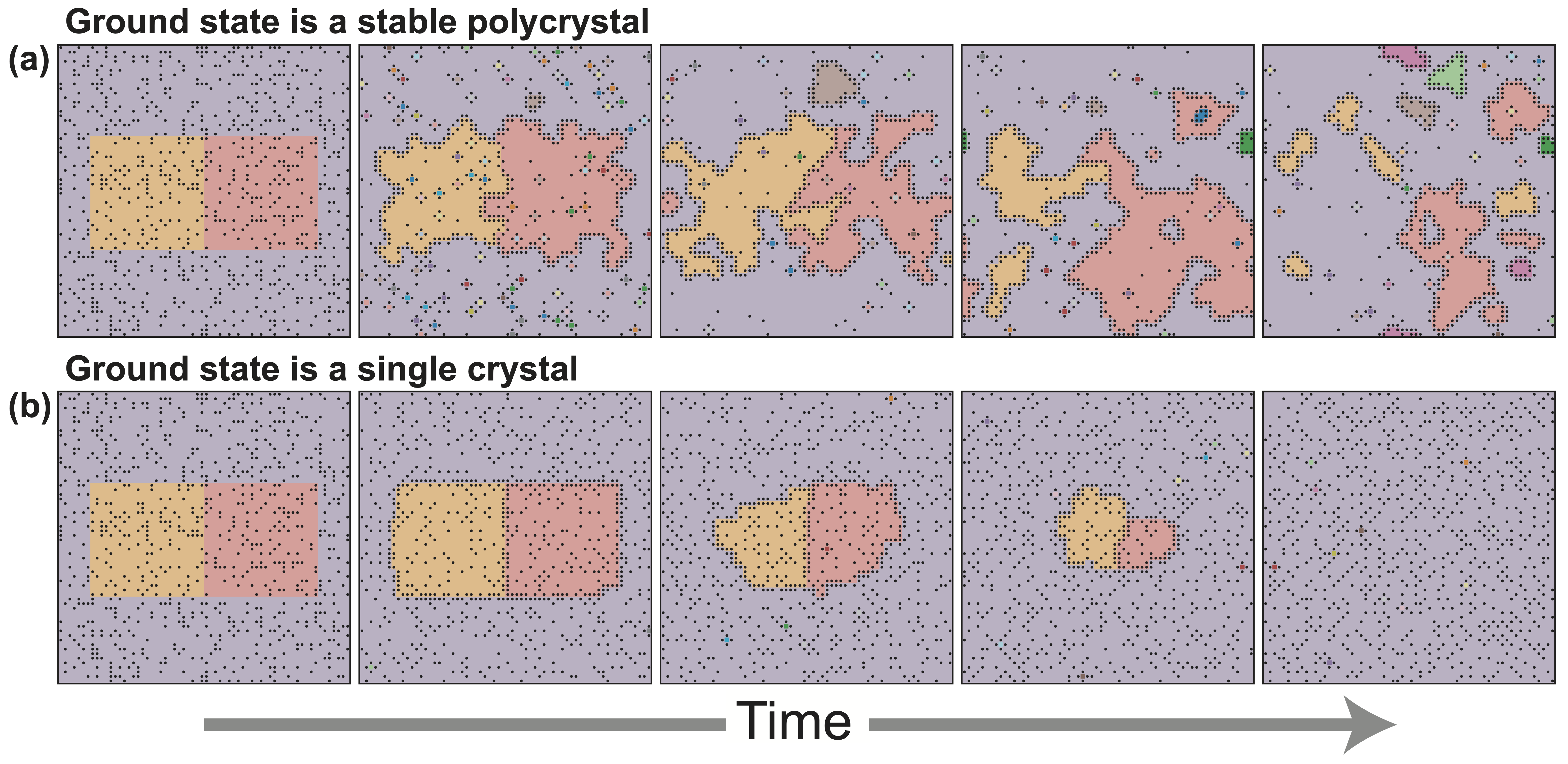}
\par\end{centering}
\caption{Time evolution (left to right) of two grains initially in contact
with each other and forming two TJs. The model parameters are $J_{sg}=-1.6$,
$J_{ss}=0.25$, $J_{ssg}=0.05$. (a) The alloy at $T=0.1$ and $c=0.2$
when the ground state is a stable polycrystal. (b) The alloys at $T=0.1$
and $c=0.2$ when the ground state is a single crystal. \label{fig:2-grains}}

\end{figure}

Table 2 summarizes the results for three representative parameter
sets corresponding to the phase diagrams shown in Figures \ref{fig:Phase-diagram-1},
\ref{fig:Phase-diagram-2}, and \ref{fig:Phase-diagram-3}. The predicted
magnitudes of $E_{s}$, $\Omega$, $\Omega^{\prime}$, and $H_{\mathrm{mix}}$
are well within the typical ranges reported in previous publications
\citep{Trelewicz2009,Chookajorn2012,Chookajorn2014,Kalidindi:2015aa,Kalidindi:2017aa,Kalidindi:2017bb,Kalidindi:2017cc}.
A notable feature of the present simulations is that the stable polycrystalline
structure was found in alloys with a \emph{negative} heat of mixing,
corresponding to the ordering trend (Table 2 and Fig.~\ref{fig:Phase-diagram-3}).
Most of the previous calculations \citep{Trelewicz2009,Chookajorn2012,Chookajorn2014,Kalidindi:2015aa,Kalidindi:2017aa,Kalidindi:2017bb,Kalidindi:2017cc}
considered phase-separating alloys with $H_{\mathrm{mix}}>0$. However,
recent publications \citep{Kalidindi:2015aa,Kalidindi:2017aa} extended
the previous analyses to alloys with $H_{\mathrm{mix}}<0$ and still
reported stable nanocrystalline states. For example, Figure 6 in \citep{Kalidindi:2015aa}
presents a diagram showing a nanocrystalline stability domain at $\Omega^{\prime}<\Omega<0$.
Furthermore, the values of $\Omega$ and $\Omega^{\prime}$ for the
representative point g in this domain are comparable to the respective
values in Table 2. A more detailed quantitative comparison with previous
publications \citep{Trelewicz2009,Chookajorn2012,Chookajorn2014,Kalidindi:2015aa,Kalidindi:2017aa,Kalidindi:2017bb,Kalidindi:2017cc}
is complicated due to intrinsic differences between models (see Supplementary
Information for more detail), but the qualitative agreement indicates
that the model parameters investigated in this paper are practically
relevant.

Due to the 2D constraint and the generic character of this model,
the results do not allow us to pinpoint a particular alloy system
in which a nano-stabilization can be achieved. However, we believe
that the results can be useful as a general guide in this pursuit.
In particular, they provide a glimpse into what a fully stabilized
NC state would look like if it were achieved experimentally in the
future.

\bigskip{}
\bigskip{}

\noindent\textbf{Acknowledgements} 

This research was supported by the National Science Foundation, Division
of Materials Research, under Award no. DMR-2103431.


\newpage{}

\begin{table}[ht]
\centering \caption{Table of physical and normalized variables in this model.}
\bigskip{}

\begin{tabular}{lcc}
\hline 
Variable & Physical & Normalized\tabularnewline
\hline 
Temperature & $T$ & $k_{B}T/J_{gg}$\tabularnewline
Total energy & $E$ & $E/J_{gg}$\tabularnewline
Solute--interface interaction energy & $J_{sg}$ & $J_{sg}/J_{gg}$\tabularnewline
Solute interaction energy in grains & $J_{ss}$ & $J_{ss}/J_{gg}$\tabularnewline
Additional solute interaction energy in GBs & $J_{ssg}$ & $J_{ssg}/J_{gg}$\tabularnewline
Time & $t$ & $t\nu_{0}$\tabularnewline
GB energy & $\gamma$ & $\gamma A/J_{gg}$\tabularnewline
\hline 
\end{tabular}\label{tab:variables}
\end{table}

\bigskip{}
\bigskip{}

\begin{table}[h]
\caption{Potts model parameters normalized to $J_{gg}$ are compared with thermodynamic
properties: grain boundary segregation energy $E_{s}$, regular
solution parameters inside the grains ($\Omega$) and in grain boundaries
($\Omega^{\prime}$), and dilute heat of mixing ($H_{\mathrm{mix}}$).
The cases 1, 2, and 3 correspond to parameter sets used for computing
the phase diagrams shown in Figures  \ref{fig:Phase-diagram-1},
\ref{fig:Phase-diagram-2}, and \ref{fig:Phase-diagram-3} of the main text.}
\label{tab:estimated-parameters}
\bigskip
\begin{centering}
\begin{tabular}{|lcrr|}
\hline 
 Parameters & Case 1 & Case 2 & Case 3\tabularnewline
\hline 
$J_{sg}/J_{gg}$ & $-1.6$ & $-1.6$ & $-1.6$\tabularnewline
\hline 
$J_{ss}/J_{gg}$ & $0$ & $-0.25$ & $0.25$\tabularnewline
\hline 
$J_{ssg}/J_{gg}$ & $0$ & $-0.05$ & $0.05$\tabularnewline
\hline 
$E_{s}$ (kJ/mol) & $-136$ & $-136$ & $-136$\tabularnewline
\hline 
$\Omega$ (kJ/mol) & $0$ & $10.61$ & $-10.61$\tabularnewline
\hline 
$\Omega^{\prime}$ (kJ/mol) & $0$ & $12.73$ & $-12.73$\tabularnewline
\hline 
$H_{\mathrm{mix}}$ (kJ/mol) & $0$ & $50.94$ & $-50.94$\tabularnewline
\hline 
\end{tabular}
\par\end{centering}

\end{table} \label{tab:estimated-parameters}

\newpage\clearpage{}
\begin{center}
\setcounter{figure}{0}\setcounter{equation}{0}\setcounter{page}{1}
\par\end{center}

\begin{center}
{\large\textbf{SUPPLEMENTARY INFORMATION}}{\large\par}
\par\end{center}

\begin{center}
{\Large\textbf{A model of full thermodynamic stabilization of nanocrystalline
alloys}}{\Large\par}
\par\end{center}

\begin{center}
{\large\bigskip{}
}{\large\par}
\par\end{center}

\begin{center}
{\large Omar Hussein and Yuri Mishin\bigskip{}
}{\large\par}
\par\end{center}

\lyxaddress{\begin{center}
{\large Department of Physics and Astronomy, MSN 3F3, George Mason
University, Fairfax, Virginia 22030, USA}
\par\end{center}}

\begin{figure}[H]
\begin{centering}
\includegraphics[width=0.7\textwidth]{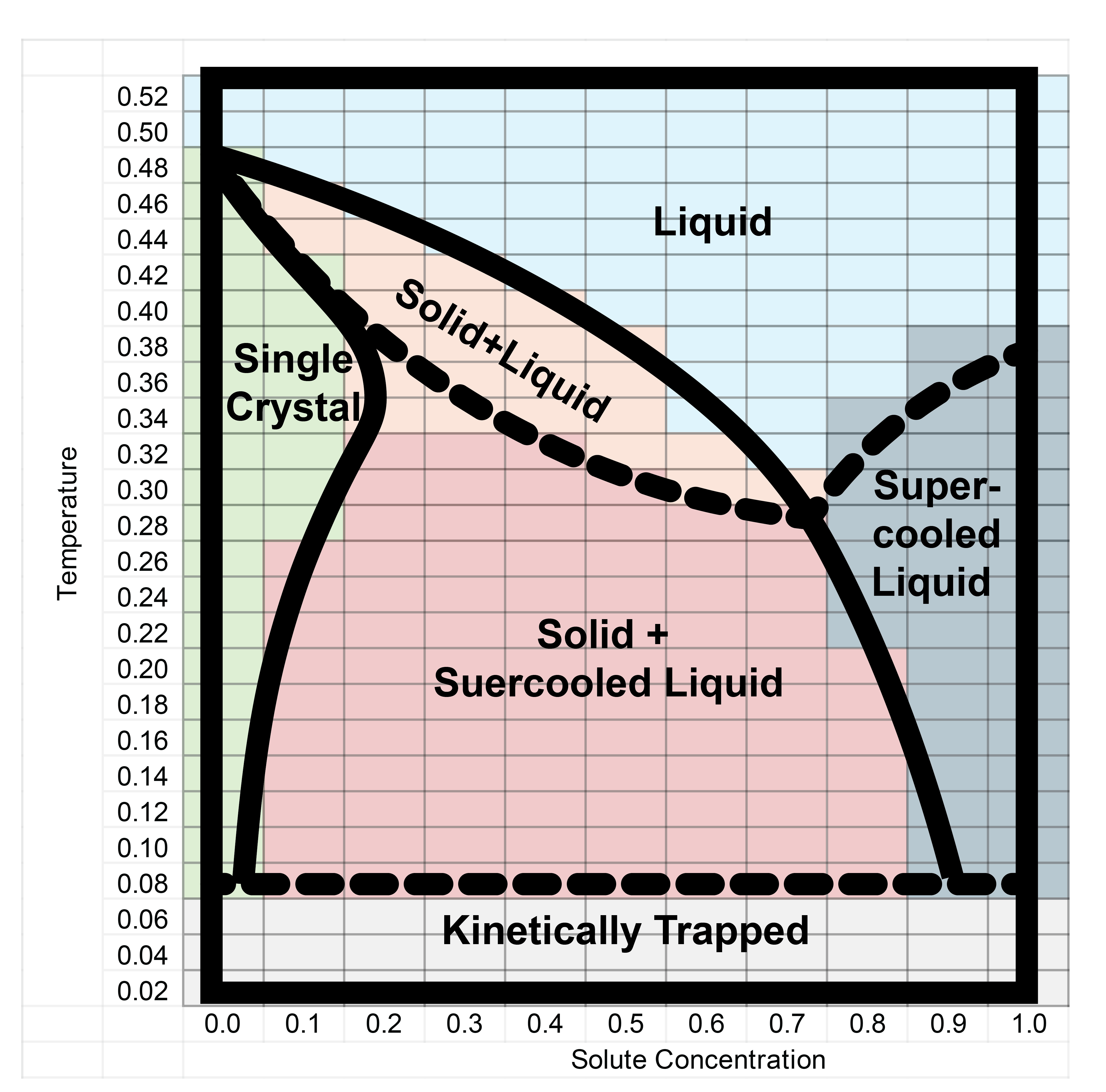}
\par\end{centering}
\caption{Example of phase diagram construction showing the grid of simulations.
The model parameters are $J_{sg}=-1.6$, $J_{ss}=J_{ssg}=0$. \label{fig:Example-of-phase}}

\end{figure}

\begin{figure}
\begin{centering}
\includegraphics[totalheight=0.82\textheight]{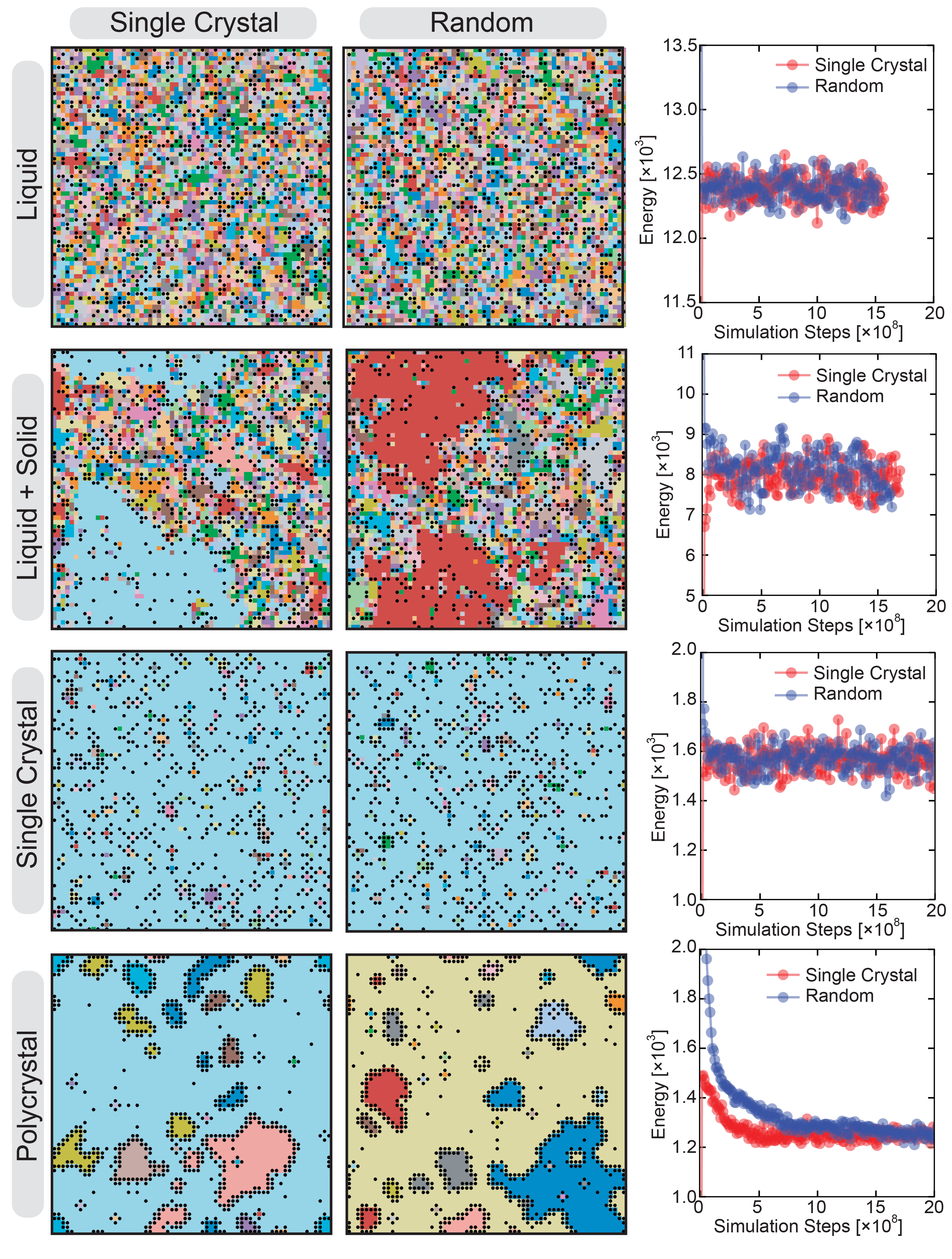}
\par\end{centering}
\caption{Representative snapshots of equilibrium configurations and energy
convergence plots for each phase on the phase diagram with positive
solute-solute interaction energies, starting from either a single
crystal or a disordered state with a random distribution of orientations.
The site orientations are color-coded and the solute atoms are represented
by black dots.\label{fig:snapshots}}
\end{figure}

\begin{figure}
\begin{centering}
\includegraphics[width=1\textwidth]{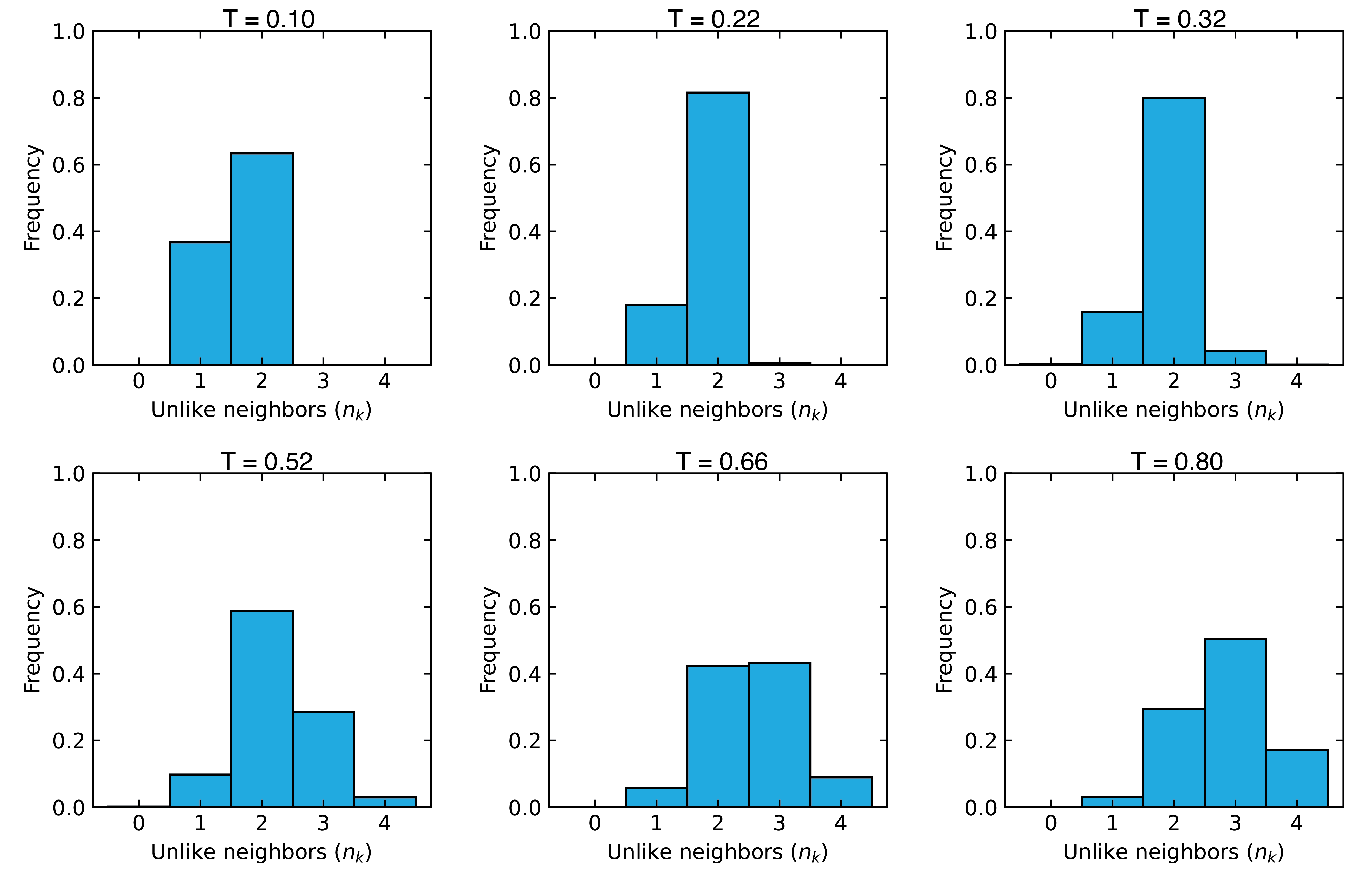}
\par\end{centering}
\caption{Selected histograms of the number of unlike neighbors $n_{k}$ at
different temperature in pure solvent liquid ($c=1$). The model parameters
are $J_{sg}=-1.6$, $J_{ss}=J_{ssg}=0$.\label{fig:histograms}}
\end{figure}


\section{Mapping of Potts model onto solution thermodynamics}

Consider a binary solid solution of components A and B with nearest-neighbor
bond energies $\varepsilon_{AA}$, $\varepsilon_{AB}$, and $\varepsilon_{BB}$.
The total energy of the solution can be written in our notation as
\begin{eqnarray}
    E_{c} & = & \dfrac{1}{2}\sum_{k}\sum_{(k,l)}\xi_{k}\left[\xi_{l}\varepsilon_{BB}+(1-\xi_{l})\varepsilon_{AB}\right]\nonumber \\
    & + & \dfrac{1}{2}\sum_{k}\sum_{(k,l)}(1-\xi_{k})\left[\xi_{l}\varepsilon_{AB}+(1-\xi_{l})\varepsilon_{AA}\right]\nonumber \\
    & = & \dfrac{1}{2}\sum_{k}\sum_{(k,l)}\varepsilon_{AA}+\sum_{k}\sum_{(k,l)}\xi_{k}(\varepsilon_{AB}-\varepsilon_{AA})\nonumber \\
    & + & \dfrac{1}{2}\sum_{k}\sum_{(k,l)}\xi_{k}\xi_{l}(\varepsilon_{BB}+\varepsilon_{AA}-2\varepsilon_{AB}).\label{eq:1-2}
\end{eqnarray}
Here, $\xi_{k}=1$ if the site (cell) $k$ is occupied by an atom
of species $B$ and $\xi_{k}=0$ otherwise. The symbol $(kl)$ indicates the
summation over all nearest-neighbors $l$ of a site $k$. 

In addition to the occupation of the solutes, the sites are characterized by orientations.
To include the orientational energy, we introduce the wrong bond parameter
$m_{kl}$ such that $m_{kl}=1$ if the neighboring sites $k$ and
$l$ have different orientations and $m_{kl}=0$ otherwise. Eq.(\ref{eq:1-2})
is modified by replacing $\varepsilon_{AA}$, $\varepsilon_{AB}$,
and $\varepsilon_{BB}$ with, respectively,
\[
\varepsilon_{AA}+m_{kl}(\varepsilon_{AA}^{\prime}-\varepsilon_{AA}),\:\:\:\varepsilon_{AB}+m_{kl}(\varepsilon_{AB}^{\prime}-\varepsilon_{AB}),\:\:\:\varepsilon_{BB}+m_{kl}(\varepsilon_{BB}^{\prime}-\varepsilon_{BB}).
\]
Here, the primed energies represent distorted bonds between misoriented
sites. Because Eq.(\ref{eq:1-2}) is linear in the bond energies,
it can be immediately rewritten in the form
\begin{eqnarray}
    E & = & E_{c}+\dfrac{1}{2}\sum_{k}\sum_{(k,l)}m_{kl}(\varepsilon_{AA}^{\prime}-\varepsilon_{AA})+\sum_{k}\sum_{(k,l)}\xi_{k}m_{kl}\left[(\varepsilon_{AB}^{\prime}-\varepsilon_{AA}^{\prime})-(\varepsilon_{AB}-\varepsilon_{AA})\right]\nonumber \\
    & + & \dfrac{1}{2}\sum_{k}\sum_{(k,l)}m_{kl}\xi_{k}\xi_{l}\left[(\varepsilon_{BB}^{\prime}+\varepsilon_{AA}^{\prime}-2\varepsilon_{AB}^{\prime})-(\varepsilon_{BB}+\varepsilon_{AA}-2\varepsilon_{AB}\right].\label{eq:2-1}
\end{eqnarray}
This equation can be simplified as follows:
\begin{eqnarray}
    E & = & E_{c}+\dfrac{1}{2}\sum_{k}(\varepsilon_{AA}^{\prime}-\varepsilon_{AA})n_{k}+\sum_{k}\left[(\varepsilon_{AB}^{\prime}-\varepsilon_{AA}^{\prime})-(\varepsilon_{AB}-\varepsilon_{AA})\right]\xi_{k}n_{k}\nonumber \\
    & + & \dfrac{1}{2}\sum_{k}\sum_{(k,l)}m_{kl}\xi_{k}\xi_{l}\left[(\varepsilon_{BB}^{\prime}+\varepsilon_{AA}^{\prime}-2\varepsilon_{AB}^{\prime})-(\varepsilon_{BB}+\varepsilon_{AA}-2\varepsilon_{AB}\right],\label{eq:2-1-1}
\end{eqnarray}
where $n_{k}=\sum_{(k,l)}m_{kl}$ is the number of unlike (wrong)
nearest-neighbors of the site $k$. 

In the Potts model with solutes introduced in the main text of the
paper, the total energy is postulated in the form
\begin{eqnarray}
    E & = & \sum_{k}J_{gg}n_{k}\nonumber \\
    & + & \sum_{k}J_{s}\xi_{k}+\sum_{k}J_{sg}\xi_{k}\phi(n_{k})\nonumber \\
    & + & \dfrac{1}{2}\sum_{k}\sum_{(kl)}\xi_{k}\xi_{l}J_{ss}+\dfrac{1}{2}\sum_{k}\sum_{(kl)}\xi_{k}\xi_{l}J_{ssg}\phi(n_{k,l}).\label{eq:4-2}
\end{eqnarray}
Here, $\phi(n_{k})$ is the grain boundary locator function defined
in the main text, $n_{k,l}$ is the average number of wrong bonds
of the sites $k$ and $l$, and $J_{s}$ is the energy on site of a solute
atom. The remaining $J$ factors in Eq.(\ref{eq:4-2}) characterize
the interaction energies between misoriented sites ($J_{gg}$), between the
solute atoms and grain boundaries ($J_{sg}$), and between solute
atoms inside the grains ($J_{ss}$) and at grain boundaries ($J_{ssg}$). 

Let us rewrite Eq.(\ref{eq:2-1-1}) in a form closest to Eq.(\ref{eq:4-2}):
\begin{eqnarray}
    E & = & \dfrac{1}{2}\sum_{k}\sum_{(k,l)}\varepsilon_{AA}+\dfrac{1}{2}\sum_{k}(\varepsilon_{AA}^{\prime}-\varepsilon_{AA})n_{k}\nonumber \\
    & + & \sum_{k}\sum_{(k,l)}\xi_{k}(\varepsilon_{AB}-\varepsilon_{AA})+\sum_{k}\left[(\varepsilon_{AB}^{\prime}-\varepsilon_{AA}^{\prime})-(\varepsilon_{AB}-\varepsilon_{AA})\right]\xi_{k}n_{k}\nonumber \\
    & + & \dfrac{1}{2}\sum_{k}\sum_{(k,l)}\xi_{k}\xi_{l}(\varepsilon_{BB}+\varepsilon_{AA}-2\varepsilon_{AB})\nonumber \\
    & + & \dfrac{1}{2}\sum_{k}\sum_{(k,l)}m_{kl}\xi_{k}\xi_{l}\left[(\varepsilon_{BB}^{\prime}+\varepsilon_{AA}^{\prime}-2\varepsilon_{AB}^{\prime})-(\varepsilon_{BB}+\varepsilon_{AA}-2\varepsilon_{AB})\right].\label{eq:2-1-1-1}
\end{eqnarray}
Our goal is to establish relationships between the $J$-factors in
Eq.(\ref{eq:4-2}) and the bond energies $\varepsilon_{ij}$ in Eq.(\ref{eq:2-1-1-1})
by mapping the two equations to each other term by term.

The first term in Eq.(\ref{eq:2-1-1-1}) is the energy of the pure
solvent A. This term is not present in Eq.(\ref{eq:4-2}). Since
this term is a constant, we can drop it without loss of generality
by assuming that $\varepsilon_{AA}=0$. The second term in Eq.(\ref{eq:2-1-1-1})
maps onto the solute-free energy:
\[
\dfrac{1}{2}\sum_{k}(\varepsilon_{AA}^{\prime}-\varepsilon_{AA})n_{k}=\sum_{k}J_{gg}n_{k},
\]
from which
\begin{equation}
    J_{gg}=\dfrac{1}{2}(\varepsilon_{AA}^{\prime}-\varepsilon_{AA}).\label{eq:3}
\end{equation}
The third term is
\begin{equation}
    \sum_{k}\sum_{(k,l)}\xi_{k}(\varepsilon_{AB}-\varepsilon_{AA})=\sum_{k}\xi_{k}z(\varepsilon_{AB}-\varepsilon_{AA}),\label{eq:3a}
\end{equation}
where $z$ is the number of closest neighbors ($z=4$ in this model).
Here, $z(\varepsilon_{AB}-\varepsilon_{AA})$ has the meaning of the
energy cost of replacing an atom A with an atom B in the pure-A perfect
lattice. In this process, we break $z$ AA bonds and create $z$ AB
bonds. The term in Eq.(\ref{eq:3a}) maps on the term $\sum_{k}J_{s}\xi_{k}$
appearing Eq.(\ref{eq:4-2}). It follows that
\begin{equation}
    J_{s}=z(\varepsilon_{AB}-\varepsilon_{AA}).\label{eq:9}
\end{equation}

The next term,
\begin{equation}
    \sum_{k}\left[(\varepsilon_{AB}^{\prime}-\varepsilon_{AA}^{\prime})-(\varepsilon_{AB}-\varepsilon_{AA})\right]\xi_{k}n_{k},\label{eq:4}
\end{equation}
represents solute interactions with grain boundaries and corresponds
to the term $\sum_{k}J_{sg}\xi_{k}\phi(n_{k})$ in Eq.(\ref{eq:4-2}).
Direct mapping in not possible because Eq.(\ref{eq:4}) represents
interactions of solute atoms with all sites that have wrong neighbors
($n_{k}\neq0$), whereas the term $\sum_{k}J_{sg}\xi_{k}\phi(n_{k})$
includes the function $\phi(n_{k})$ specifically targeting grain
boundary environments. We can only say that $J_{sg}$ is physically
related to the energy difference $(\varepsilon_{AA}^{\prime}-\varepsilon_{AB}^{\prime})-(\varepsilon_{AA}-\varepsilon_{AB})$. 

The term
\[
\dfrac{1}{2}\sum_{k}\sum_{(k,l)}\xi_{k}\xi_{l}(\varepsilon_{BB}+\varepsilon_{AA}-2\varepsilon_{AB}),
\]
describes solute-solute interactions and is similar to the term $\tfrac{1}{2}\sum_{k}\sum_{(kl)}\xi_{k}\xi_{l}J_{ss}$
in Eq.(\ref{eq:4-2}). This leads to the mapping 
\begin{equation}
    J_{ss}=(\varepsilon_{BB}+\varepsilon_{AA}-2\varepsilon_{AB}).\label{eq:4-1}
\end{equation}
The last term in Eq.(\ref{eq:2-1-1-1}), 
\begin{equation}
    \dfrac{1}{2}\sum_{k}\sum_{(k,l)}m_{kl}\xi_{k}\xi_{l}\left[(\varepsilon_{BB}^{\prime}+\varepsilon_{AA}^{\prime}-2\varepsilon_{AB}^{\prime})-(\varepsilon_{BB}+\varepsilon_{AA}-2\varepsilon_{AB})\right],\label{eq:5}
\end{equation}
is similar to the term $\tfrac{1}{2}\sum_{k}\sum_{(kl)}\xi_{k}\xi_{l}J_{ssg}\phi(n_{k,l})$
in Eq.(\ref{eq:4-2}). However, direct mapping is again impossible.
Eq.(\ref{eq:5}) accounts for the difference in solute-solute interactions
inside the perfect lattice and between solute atoms occupying sites
with wrong neighbors, whereas the Potts model makes this effect specific
to grain boundaries. Furthermore, the Potts model considers the average
``wrongness'' $n_{k,l}$ of the two sites. All we can say is that
the parameter $J_{ssg}$ is physically related to the change in the
interaction parameter $(\varepsilon_{BB}+\varepsilon_{AA}-2\varepsilon_{AB})$
when the solute is at grain boundaries.

It is instructive to reformulate the above equations in terms of the
regular solution parameter
\begin{equation}
    \Omega=\varepsilon_{AB}-\dfrac{\varepsilon_{BB}+\varepsilon_{AA}}{2}.\label{eq:7}
\end{equation}
The solute-solute interaction factor (\ref{eq:4-1}) becomes 
\begin{equation}
    J_{ss}=-2\Omega,\label{eq:8}
\end{equation}
while Eq.(\ref{eq:5}) takes the form
\begin{equation}
    -\sum_{k}\sum_{(k,l)}m_{kl}\xi_{k}\xi_{l}(\Omega^{\prime}-\Omega).\label{eq:5-1}
\end{equation}

\begin{figure}[h]
    \begin{centering}
        \includegraphics[width=0.4\textwidth]{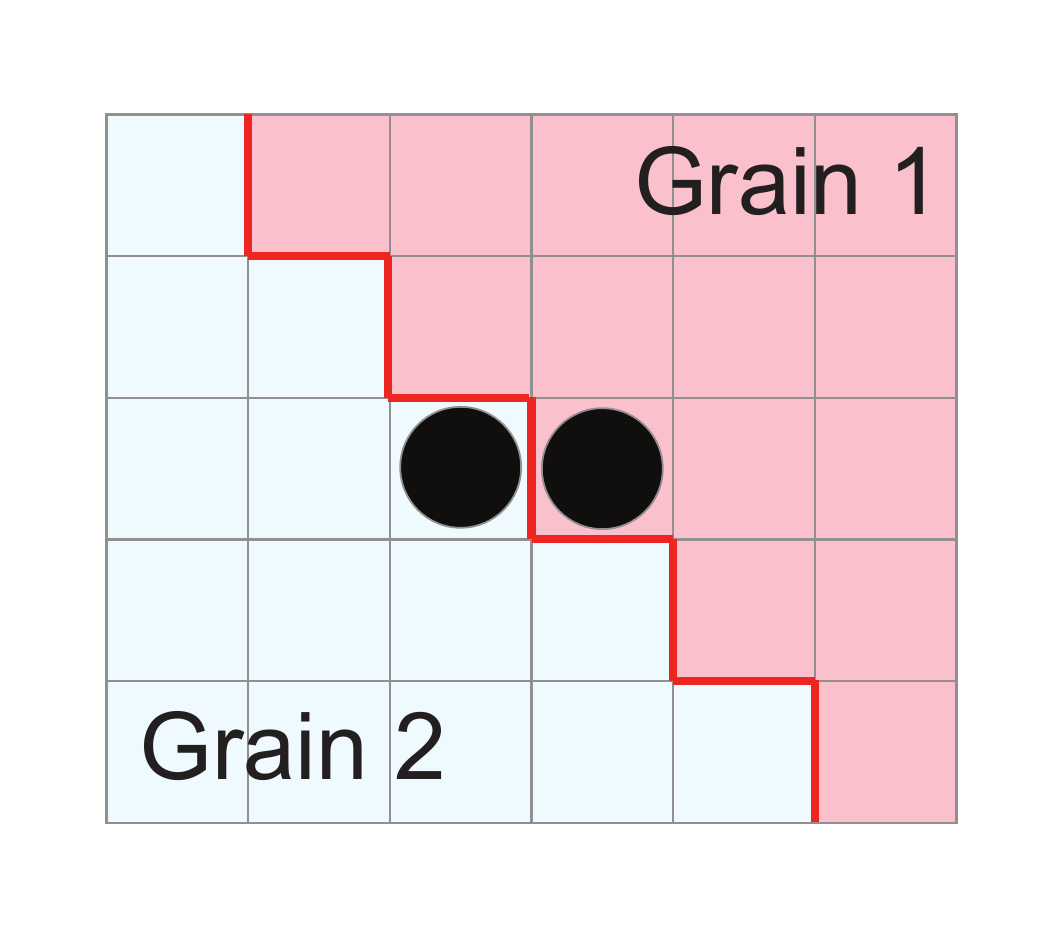}
        \par\end{centering}
    \caption{Schematic grain boundary separating perfectly crystalline grains.
        The solid red segments represent boundaries between sites with different
        orientations. The two black circles symbolize solute atoms segregated
        at the grain boundary.\label{fig:grain-boundary}}
\end{figure}

\section{Application to a single planar grain boundary}

In this section, we apply the general mapping relations established
in the previous section to a particular case of a single planar grain
boundary shown schematically in Figure \ref{fig:grain-boundary}.
The grains are considered perfectly single-crystalline ($n_{k}=0$).
In addition, we will specialize the above equations in the constraint
$J_{s}=0$, which we adopted in this work to simplify the parametric
study. Eq.(\ref{eq:9}) gives $\varepsilon_{AB}-\varepsilon_{AA}=0$.
Considering that we have shifted the reference energy to achieve $\varepsilon_{AA}=0$,
we arrive at $\varepsilon_{AB}=\varepsilon_{AA}=0$. Thus, there is
only one independent bond energy: $\varepsilon_{BB}$.

For the grain boundary geometry shown in Figure \ref{fig:grain-boundary},
all boundary sites have $n_{k}=2$ and thus $\phi(n_{k})=1$. Each
boundary site has excess energy $J_{gg}$ and occupies the grain
boundary area $A=\sqrt{2}a^{2}$, where $a$ is the side of the square
that represents the site. As a result, the grain boundary energy per
unit area in a solute-free system is
\begin{equation}
    \gamma=\dfrac{J_{gg}}{A}=\dfrac{J_{gg}}{\sqrt{2}a^{2}}.\label{eq:10}
\end{equation}

Eq.(\ref{eq:4}) predicts that the energy change in moving a single
solute atom from inside a grain to the grain boundary is 
\begin{equation}
    E_s = 2(\varepsilon_{AB}^{\prime}-\varepsilon_{AA}^{\prime})-2(\varepsilon_{AB}-\varepsilon_{AA})=2(\varepsilon_{AB}^{\prime}-\varepsilon_{AA}^{\prime}),\label{eq:4-3}
\end{equation}
whereas the respective term $\sum_{k}J_{sg}\xi_{k}\phi(n_{k})$ in
Eq.(\ref{eq:4-2}) reduces to $J_{sg}$. This allows us to identify
$J_{sg}$ as the grain boundary segregation energy:
\begin{equation}
    E_{s}=J_{sg}.\label{eq:11}
\end{equation}
According to Eq.(\ref{eq:8}), 
\begin{equation}
    \Omega=-\dfrac{J_{ss}}{2}.\label{eq:12}
\end{equation}
To identify $J_{ssg}$, we note that the term $\dfrac{1}{2}\sum_{k}\sum_{(kl)}\xi_{k}\xi_{l}J_{ssg}\phi(n_{k,l})$
in Eq.(\ref{eq:4-2}) simplifies to $\tfrac{1}{2}\sum_{k}\sum_{(kl)}\xi_{k}\xi_{l}J_{ssg}$,
while Eq.(\ref{eq:5-1}) becomes 

\begin{equation}
    -\sum_{k}\sum_{(k,l)}\xi_{k}\xi_{l}(\Omega^{\prime}-\Omega).\label{eq:5-1-1}
\end{equation}
It follows that 
\begin{equation}
    (\Omega^{\prime}-\Omega)=-\dfrac{1}{2}J_{ssg}.\label{eq:13}
\end{equation}

Equations (\ref{eq:10}), (\ref{eq:11}), (\ref{eq:12}), and (\ref{eq:13})
express the experimentally relevant properties $\gamma$, $E_{s}$,
$\Omega$, and $\Omega^{\prime}$ through model parameters $J_{gg},$
$J_{sg}$, $J_{ss}$, and $J_{ssg}$, respectively.

\section{Numerical estimates}

In the KMC simulations reported in the paper, we used the dimensionless
variables by normalizing the temperature and $J$-factors by $J_{gg}$,
see Table 1 in the main text. Inverting Eq.(\ref{eq:10}), we obtain
\begin{equation}
    J_{gg}=\gamma\sqrt{2}a^{2},\label{eq:10a}
\end{equation}
indicating that $J_{gg}$ can be interpreted as the energy per atom in
the grain boundary core. 

For numerical estimates, we will use the typical grain boundary energy
$\gamma=1\:\mathrm{J}/\mathrm{m}^{2}$ and estimate $a$ by 1 nm.
Eq.(\ref{eq:10a}) then gives us $J_{gg}=0.88$ eV/atom, which is
a physically reasonable number for the excess energy of a grain boundary
atom. Using this value of $J_{gg}$, we calculated the energies $J_{sg}$,
$J_{ss}$, and $J_{ssg}$ and then obtained the thermodynamic properties
$E_{s}$, $\Omega$, and $\Omega^{\prime}$ for the three cases for
which the phase diagrams were computed in the paper. 

\begin{table}
    \begin{centering}
        \begin{tabular}{|l|c|r|r|}
            \hline 
            & Case 1 & Case 2 & Case 3\tabularnewline
            \hline 
            \hline 
            $J_{sg}/J_{gg}$ & $-1.6$ & $-1.6$ & $-1.6$\tabularnewline
            \hline 
            $J_{ss}/J_{gg}$ & $0$ & $-0.25$ & $0.25$\tabularnewline
            \hline 
            $J_{ssg}/J_{gg}$ & $0$ & $-0.05$ & $0.05$\tabularnewline
            \hline 
            $E_{s}$ (eV/atom) & $-1.41$ & $-1.41$ & $-1.41$\tabularnewline
            \hline 
            $E_{s}$ (kJ/mol) & $-136$ & $-136$ & $-136$\tabularnewline
            \hline 
            $\Omega$ (eV/atom) & $0$ & $0.110$ & $-0.110$\tabularnewline
            \hline 
            $\Omega$ (kJ/mol) & $0$ & $10.61$ & $-10.61$\tabularnewline
            \hline 
            $\Omega^{\prime}$ (eV/atom) & $0$ & $0.132$ & $-0.132$\tabularnewline
            \hline 
            $\Omega^{\prime}$ (kJ/mol) & $0$ & $12.73$ & $-12.73$\tabularnewline
            \hline 
            $H_{\mathrm{mix}}$ (kJ/mol) & $0$ & $50.94$ & $-50.94$\tabularnewline
            \hline 
        \end{tabular}
        \par\end{centering}
    \caption{Potts model parameters normalized to $J_{gg}$ are compared with thermodynamic
        properties: grain boundary segregation energy $E_{s}$, the regular
        solution parameters inside the grains ($\Omega$) and in grain boundaries
        ($\Omega^{\prime}$), and the dilute heat of mixing ($H_{\mathrm{mix}}$).
        The cases 1, 2, and 3 correspond to parameter sets used for computing
        the phase diagrams shown in Figures 3, 6 and 7 of the main text.\label{tab:estimated-parameters}}
    
\end{table}

Table \ref{tab:estimated-parameters} summarizes the results in both
eV/atom and practical units kJ/mol. In addition to the regular solution
parameter $\Omega$, the table includes the perfect-lattice heat of
mixing $H_{\mathrm{mix}}=z\Omega$ in the dilute limit. 

The predicted magnitudes of $E_{s}$, $\Omega$, $\Omega^{\prime}$,
and $H_{\mathrm{mix}}$ are well within the range of typical values
reported in previous publications \citep{Trelewicz2009,Chookajorn2012,Murdoch:2013ab,Chookajorn2014,Kalidindi:2015aa,Kalidindi:2017cc}.
The notable feature of the present simulations is that the stable
polycrystalline structure was found in alloys with a negative heat
of mixing, corresponding to the ordering trend (Case 3 in Table \ref{tab:estimated-parameters}).
Most of the previous calculations \citep{Trelewicz2009,Chookajorn2012,Murdoch:2013ab,Chookajorn2014,Kalidindi:2015aa,Kalidindi:2017cc}
considered phase-separating alloys with $H_{\mathrm{mix}}>0$. However,
recent publications \citep{Kalidindi:2015aa,Kalidindi:2017aa} extended the previous
analyses to alloys with $H_{\mathrm{mix}}<0$, and stable nanocrystalline
states were also observed. For example, Figure 6 in~\citep{Kalidindi:2015aa}
presents a diagram showing a nanocrystalline stability domain at $\Omega^{\prime}<\Omega<0$.
Furthermore, the values of $\Omega$ and $\Omega^{\prime}$ of the representative
point g in this domain are comparable to the values for Case 3
in Table \ref{tab:estimated-parameters}. Finally, the temperature
interval of the stable polycrystalline state in Case 3 is approximately
between 1000 and 3000 K. Although these temperatures are higher than
one could expect, they are not unreasonable given the assumptions
and approximations underlying this calculation. 

A more detailed comparison with previous publications \citep{Trelewicz2009,Chookajorn2012,Murdoch:2013ab,Chookajorn2014,Kalidindi:2015aa,Kalidindi:2017cc}
is complicated due to intrinsic differences between models.
In addition, in both models, different constraints are imposed to
reduce the dimensionality of the parameter space. For example, we
imposed the constraint $\varepsilon_{AB}=\varepsilon_{AA}=0$, whereas
the previous models \citep{Trelewicz2009,Chookajorn2012,Murdoch:2013ab,Chookajorn2014,Kalidindi:2015aa,Kalidindi:2017cc}
imposes the constraints (in our notation) $\varepsilon_{AA}=\varepsilon_{BB}$
and $\varepsilon_{AA}^{\prime}=\varepsilon_{BB}^{\prime}$.


\end{document}